\documentclass[useAMS,usegraphicx,usenatbib]{mn2e}

\usepackage{amssymb}
\usepackage{times}

\title[Classical novae from POINT--AGAPE -- I.]{Classical novae from
the POINT-AGAPE microlensing survey of M31 -- I. The
nova catalogue\thanks{Based on observations made with the Isaac Newton Telescope
operated on the island of La Palma by the Isaac Newton Group in the
Spanish Observatorio del Roque de los Muchachos of the Instituto
de Astrofisica de Canarias.}}

\author[M.J.~Darnley et al.]  {M.J.~Darnley$^1$\thanks{E-mail:
mjd@astro.livjm.ac.uk}, M.F.~Bode$^1$, E.~Kerins$^1$, A.M.~Newsam$^1$,
J.~An$^2$, P.~Baillon$^3$,\newauthor 
S.~Calchi~Novati$^4$, B.J.~Carr$^5$, M.~Cr\'ez\'e$^{6,7}$, N.W.~Evans$^2$, 
Y.~Giraud-H\'eraud$^6$,\newauthor
A.~Gould$^8$, P.~Hewett$^2$, Ph.~Jetzer$^4$, 
J.~Kaplan$^6$,S.~Paulin-Henriksson$^6$,\newauthor 
S.J.~Smartt$^2$, C.S.~Stalin$^6$ and Y.~Tsapras$^5$\\
$^{1}$Astrophysics Research 
Institute, Liverpool John Moores University, 12 Quays House, Egerton Wharf, 
Birkenhead, CH41 1LD, UK\\
$^2$Institute of Astronomy, University of Cambridge, 
Madingley Road, Cambridge CB3 0HA, UK\\
$^3$CERN, CH-1211 Gen\`eve 23, 
Switzerland\\
$^4$Institut f\"ur Theoretische Physik, Universit\"at Z\"urich,
Winterthurerstrasse 190, CH-8057 Z\"urich, Switzerland\\
$^5$Astronomy Unit, School of Mathematical Sciences, Queen Mary, University of
London, Mile End Road, London E1 4NS, UK\\
$^6$Laboratoire de Physique Corpusculaire et Cosmologie, UMR 7553, CNRS-IN2P3
Coll\`ege de France, 11 Place Marcelin Berthelot, F-75231 Paris, France\\
$^7$Universit\'e Bretagne-Sud, Campus de Tohannic,BP~573, F-56017 Vannes Cedex,
France\\
$^8$Department of Astronomy, Ohio State University, 140 West 18th Avenue,
Columbus, OH 43210, USA}

\begin{document}
\maketitle

\begin{abstract}
The POINT-AGAPE survey is an optical search for gravitational
microlensing events towards the Andromeda Galaxy (M31).  As well as
microlensing, the survey is sensitive to many different classes of
variable stars and transients. Here we describe the automated
detection and selection pipeline used to identify M31 classical novae
(CNe) and we present the resulting catalogue of 20 CN candidates
observed over three seasons. CNe are observed both in the bulge region
as well as over a wide area of the M31 disk. Nine of the CNe are
caught during the final rise phase and all are well sampled in at
least two colours. The excellent light-curve coverage has allowed us
to detect and classify CNe over a wide range of speed class, from very
fast to very slow. Among the light-curves is a moderately fast CN
exhibiting entry into a deep transition minimum, followed by its final
decline. We have also observed in detail a very slow CN which faded by
only 0.01~mag~day$^{-1}$ over a 150 day period. We detect other
interesting variable objects, including one of the longest period and
most luminous Mira variables. The CN catalogue constitutes a uniquely
well-sampled and objectively-selected data set with which to study the
statistical properties of classical novae in M31, such as the global
nova rate, the reliability of novae as standard-candle distance
indicators and the dependence of the nova population on stellar
environment. The findings of this statistical study will be reported
in a follow-up paper.
\end{abstract}

\begin{keywords}
novae, cataclysmic variables -- galaxies: individual: M31
\end{keywords}

\section{Introduction}

\begin{table*}
\begin{minipage}{114mm}
\caption{Principal M31 classical nova surveys}
\label{novasurveys}
\begin{tabular}{lllllll}
\hline
Author(s)              & Epoch      & Filter(s)        & Detector & Novae & 
Annual          & Reference(s) \\                       &            &           
       &          &       & rate            &              \\ \hline
Hubble                 & 1909--1927 & B                & Plates   &  85   & 
$\sim 30$       & 1       \\Arp                    & 1953--1954 & B
  & Plates   &  30   & $24 \pm 4$      & 2       \\Rosino {\it et al.}    &
1955--1986 & B                & Plates   & 142   & -               & 3, 4, 5
\\Ciardullo {\it et al.} & 1982--1986 & B, H$\alpha$     & CCD      &  40   & -
             & 6, 7    \\Sharov \& Alksins      & 1969--1989 & B
& Plates   &  21   & -               & 8       \\Tomaney \& Shafter     &
1987--1989 & H$\alpha$        & CCD      &   9   & -               & 9
\\Shafter \& Irby        & 1990--1997 & H$\alpha$        & CCD      &  72   &
$37_{-8}^{+12}$ & 10      \\Rector {\it et al.}    & 1995--1999 & H$\alpha$      
  & CCD      &  44   & -               & 11      \\Darnley {\it et al.}   &
1999--2002 & $r'$, $i'$, $g'$ & CCD      &  20   & $*$             &
12 \\
\hline
\end{tabular}
References: (1)~\citet{1929ApJ....69..103H}; 
(2)~\citet{1956AJ.....61...15A}; (3)~\citet{1964AnAp...27..498R}; 
(4)~\citet{1973A&AS....9..347R}; (5)~\citet{1989AJ.....97...83R}; 
(6)~\citet{1987ApJ...318..520C}; (7)~\citet{1990ApJ...356..472C}; 
(8)~\citet{1991Ap&SS.180..273S}; (9)~\citet{1992ApJS...81..683T}; 
(10)~\citet{2001ApJ...563..749S}; (11)~\citet{1999AAS...195.3608R}; 
(12)~This work.
\newline
$^*$ To be reported in a follow-up paper.
\end{minipage}
\end{table*}

Classical novae (CNe) undergo unpredictable outbursts with a total
energy that is surpassed only by gamma-ray bursts, supernovae
and some luminous blue variables.  However, CNe are far more commonplace
than these other phenomena \citep{1989clno.book....1W}.

CNe are a sub-class of cataclysmic variables (CVs).  The canonical
model for CVs \citep{1956ApJ...123...44C} is that they are close
binary systems, generally of short period, containing a low-mass G or
K type near-main-sequence late-type dwarf that fills its Roche lobe
(the secondary) and a more massive carbon-oxygen or magnesium-neon
white dwarf companion (the primary).  As the secondary fills its lobe,
any tendency for it to increase in size through evolutionary processes
causes a flow of material through the inner Lagrangian point into the
primary's lobe.  The high angular momentum of the transferred material
causes it to form a disc around the white dwarf.  Viscous forces
within this accretion disc act to transfer material inwards, so that a
small amount of the accreted hydrogen-rich material falls on to the
primary's surface.  As this layer grows, the temperature of
the material increases.  Eventually the temperature may become high
enough to initiate hydrogen burning.  Given the correct conditions,
this can lead to a thermonuclear runaway in which the accreted
material (and possibly some of the ``dredged-up'' white dwarf) is
expelled from the system in a nova eruption
\citep{1989clno.book...17K,1989clno.book...39S}.

CNe exhibit outburst amplitudes of $\sim 10-20$ magnitudes and, at
maximum light, display an average absolute blue magnitude of
$M_{B}=-8$ with a limit of around $M_{B}=-9.5$ for the fastest
\citep{1981ApJ...243..268S,1989clno.book....1W}.  They are potentially
useful as standard candles for extragalactic distance indication, as
they exhibit a correlation between their luminosity at maximum light
and the rate of their decline
\citep{1929ApJ....69..103H,1945PASP...57...69M,
1956AJ.....61...15A}. Unfortunately, poor light-curve coverage, small
sample sizes and a current lack of understanding of how the properties
of CNe vary between different stellar populations, have limited their
usefulness in this context.  Due to their relatively high frequency,
novae may also be used as a tool for mapping the spatial distribution
of the population of close binary systems in nearby galaxies. Despite
the high luminosities of CNe, our position within the Galaxy prevents
us from directly observing more than a small fraction of Galactic CNe
that erupt each year \citep{1997ApJ...487..226S}. Therefore the
Galactic nova rate is poorly known, with estimates ranging from
11~yr$^{-1}$
\citep{1990AJ.....99.1079C} to 260~yr$^{-1}$\citep{1972SvA....16...41S}.  
Fortunately, CNe can be readily identified in external
galaxies. Surveys of CNe in M31 have been carried out by
\citet{1929ApJ....69..103H},
\citet{1956AJ.....61...15A},
\citet{1964AnAp...27..498R,1973A&AS....9..347R}, \citet{1987ApJ...318..520C},
\citet{1989AJ.....97.1622C}, \citet{1991Ap&SS.180..273S},
\citet{1992ApJS...81..683T}, \citet{1999AAS...195.3608R} and
\citet{2001ApJ...563..749S} amongst others.  These surveys have resulted in the
discovery of around 450 novae and have indicated the global nova rate
to be $\sim 30-40$~yr$^{-1}$ \citep{2001ApJ...563..749S}.
Table~\ref{novasurveys} summarises the findings of many of these past
surveys, with most of the data reproduced from
\citet{2001ApJ...563..749S}.  The relatively high nova rate of M31 and
its close proximity to our own Galaxy are major advantages of
targeting M31 for nova surveys.  However, since M31 is nearer edge-on
than face-on, the task of distinguishing between possible separate
disk and bulge nova populations is difficult, and there remains debate
surrounding the distribution and rate of novae within M31.
POINT-AGAPE (Pixel-lensing Observations with the Isaac Newton
Telescope - Andromeda Galaxy Amplified Pixels Experiment) is searching
for gravitational microlensing events against the mostly unresolved
stars in M31 \citep{2003A&A...405...15P}.  It uses the wide-field
camera (WFC) on the Isaac Newton Telescope (INT) to survey a
0.6~deg$^2$ region of M31 in at least two broadband filters. The
survey has very good temporal sampling over the M31 observing season
(August -- January) for three seasons and is therefore an excellent
database within which to look for novae. Whilst H$\alpha$ observations
have been shown to be a particularly helpful diagnostic for CN
identification \citep{1957QB841.G24......,1983ApJ...272...92C}, the
excellent sampling of our survey more than makes up for an absence of
H$\alpha$ data as it allows us to classify the light-curve profiles of
different novae.

A commonly used method of describing the overall timescale of an
eruption and classifying CNe, is the concept of the nova ``speed
class'', developed by \citet{1939PA.....47..410M} and \citet{1948AnAp...11....3B}.  Their definition of
the various nova classes depends on the time taken for a nova to
diminish by three magnitudes below maximum light, $t_{3}$.  Throughout this paper we will use the speed-class
definition, modified by \citet{1957QB841.G24......} for $t_{2}$ times, given by \citet{1989clno.book....1W},
reproduced in Table~\ref{speed}. The 2.5-year baseline of our survey,
along with the high sampling rate, allows us to classify novae over a
wide range of speed classes.

\begin{table}
\caption{The classification of classical nova light-curves into
various speed classes according to the time taken to decrease in
brightness by two magnitudes $(t_{2})$ and their $V$-band rate of
decline $(dV/dt)$ from maximum light
\protect\citep{1989clno.book....1W}.}
\label{speed}
\begin{tabular}{lll}
\hline
Speed class     & $t_{2}$  & $dV/dt$\\
                & (days)   & (mag day$^{-1}$)\\
\hline
Very fast       & $\leq 10$    & $\geq 0.20$ \\
Fast            & 11--25   & 0.18--0.08 \\
Moderately Fast & 26--80   & 0.07--0.025 \\
Slow            & 81--150  & 0.024--0.013\\
Very slow       & 151--250 & 0.013--0.008 \\
\hline
\end{tabular}
\end{table}

In order for robust statistical statements to be made about CN
populations in external galaxies it is important to take account of
the bias induced by the unresolved galactic surface brightness
component, which in the inner regions of galaxies may mask much of the
CNe light-curve evolution, making identification and classification
more difficult. One approach to this is to make the selection
procedure for CN candidates completely automated
\citep{2002cne..conf..481D}, so that the detection efficiency for the
various CN light-curve morphologies may be assessed as a function of
position objectively through Monte Carlo simulation. This is a far
from trivial task because the light-curve structure of CNe varies
considerably with nova speed class. Automation of the pipeline also
requires that, in the absence of H$\alpha$ observations, the
light-curves are well sampled.

The aim of this study is to present the first fully automated search
for CN which can be used to assess in an objective manner the
statistical properties of the CN population in M31. The study includes
some of the finest examples of extragalactic nova light-curves
observed to date. In a companion work we will study the spatial
distribution and rate of novae in M31.  We will also assess the
potential of our nova candidates as distance indicators by calibrating the
maximum-magnitude versus rate-of-decline and and other proposed 
relationships \citep{1955Obs....75..170B}.

The outline of this paper is as follows. In Section~\ref{dataset} we
describe the POINT-AGAPE survey dataset. In Section~\ref{reduction} we
detail the initial data reduction stages. Section~\ref{detectpipe}
describes the main nova detection pipeline used to define our CN
catalogue. The catalogue itself is presented in
Section~\ref{catalogue} and we discuss our main findings in 
Section~\ref{conclusion}.

\section{The POINT-AGAPE dataset} \label{dataset}

\begin{figure}
\rotatebox{270}{\includegraphics[width=243pt]{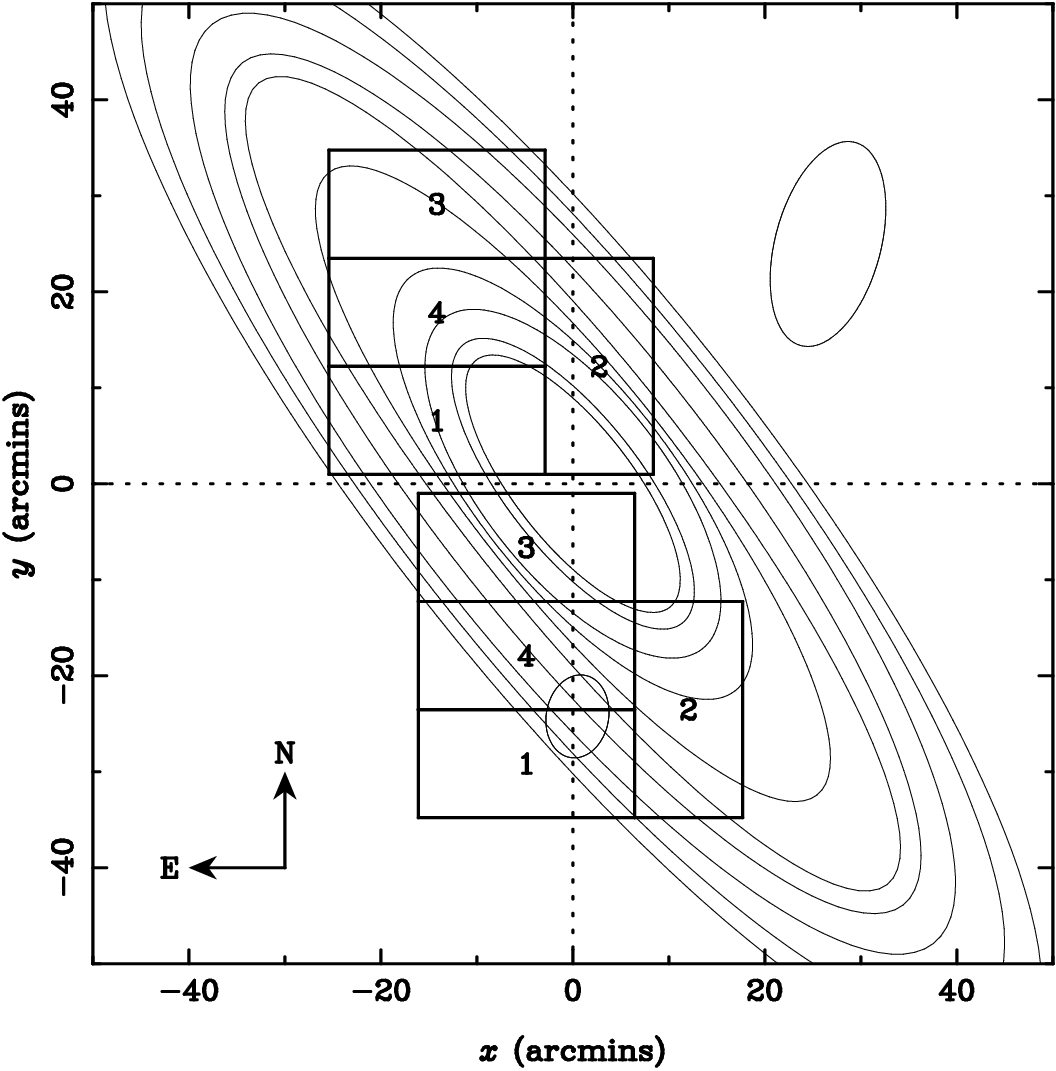}}
\caption{The positions of the North and South POINT-AGAPE fields.
Each rectangle represents one of the four WFC CCDs and are numbered
accordingly.  The origin is the centre of M31 at (J2000) $\alpha =
0^{\rm h}42^{\rm m}44\fs324$, $\delta = +41\degr16\arcmin08\farcs53$
\protect\citep{1992ApJ...390L...9C}. Also indicated are ten
representative M31 ``isophotes'' from the surface photometry of
\protect\citet{1958ApJ...128..465D}, along with representations of the
positions and sizes of M32 (within the southern field) and NGC205.}
\label{fields}
\end{figure}

Between the end of 1999 August and the end of 2002 January, we have
used the WFC on the INT in La~Palma to regularly monitor two fields
positioned north and south of the M31 centre and covering
0.6~deg$^{2}$. The north field is located at $\alpha = 0^{\rm
h}44^{\rm m}00\fs0$, $\delta = +41\degr34\arcmin00\farcs0$ and the
south field at $\alpha = 0^{\rm h}43^{\rm m}23\fs0$, $\delta =
+40\degr58\arcmin15\farcs0$ (J2000), with respect to the centre of CCD4.  The
WFC consists of a mosaic of
four $2048\times4100$ pixel CCDs, and the field locations are
indicated in Figure~\ref{fields}. The field placements were primarily
chosen to be sensitive to compact dark matter candidates, or Machos,
which are predicted to be most evident towards the far side of the M31
disk \citep{2001MNRAS.323...13K}.

\begin{figure}
\rotatebox{270}{\includegraphics[width=233pt]{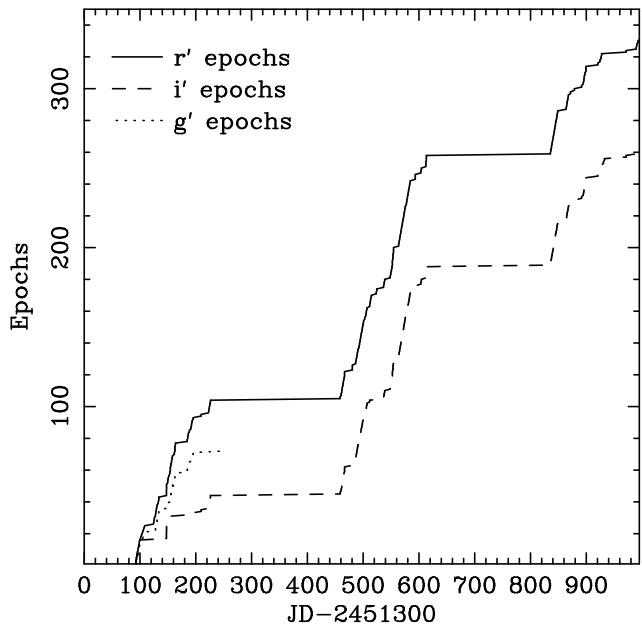}}
\caption{The cumulative temporal coverage of the POINT-AGAPE
survey in the $r'$, $i'$ and $g'$ bands for the northern field. The
Southern-field data-set has a similar distribution.}
\label{tempcov}
\end{figure}

The observations were conducted over three seasons in at least two
broad-band Sloan-like pass-bands (usually $r'$ and $i'$, though sometimes
augmented by $g'$).  The three seasons of $r'$ data comprise 333 epochs
from the northern field and 318 epochs from the
southern field.  The full distribution of observations, in each band,
over the three seasons can be seen in Table~\ref{obsdist}, whilst a
graphical representation of the temporal coverage of the
northern-field data may be seen in Figure~\ref{tempcov}.

\begin{table}
\caption{The distribution of each of the three seasons of observations
by field and filter band.}
\label{obsdist}
\begin{tabular}{llll}
\hline
Season & $r'$-band    & $i'$-band    & $g'$-band \\
       & observations & observations & observations \\
       & North/South  & North/South  & North/South \\
\hline
1      & 105/98       & 44/36        & 72/75 \\
2      & 154/150      & 144/134      & 0/0 \\
3      & 74/70        & 78/73        & 0/0 \\
\hline
\end{tabular}
\end{table}

\section{Data reduction} \label{reduction}

\subsection{Image pre-processing and alignment}

The data pre-processing is performed using the WFC reduction package
{\tt WFCRED}, the processing stages of which include:

\begin{enumerate}
{\item linearity correction;}   {\item CCD processing -- including
de-biasing and flat fielding;}   {\item de-fringing -- for $i'$ images
it is necessary to subtract a mean ``fringe frame'' to correct for
fringing effects; and} 
{\item world coordinate system definition.}
\end{enumerate}

After pre-processing, the data reduction steps below are carried out by
automated scripts from within the NOAO IRAF package environment\footnote{IRAF is
distributed by the National Optical Astronomy Observatories, which are operated
by the Association of Universities for Research in Astronomy, Inc.  under
cooperative agreement with the National Science Foundation.}.

The first step involves geometrically aligning the image stack.  This
is carried out using three packages: {\tt xyxymatch} to produce lists 
of matched reference coordinates; {\tt geomap} to calculate
second-order geometric transformations between images; and {\tt
geotran} to apply the geometric transforms to the images.
The aligned images are then trimmed to produce a common overlap
region.  The data loss due to the trimming process is summarised in
Table~\ref{lostpix}.  Overall it accounts for a loss of 4.5\% of the
initial data.

\subsection{PSF matching and background subtraction}

The next stage in the reduction process is to match the point-spread
functions (PSFs) of the $r'$ images. The data reduction and candidate
selection is performed initially in $r'$ as this dataset has the best
temporal coverage. The {\tt psfmeasure} package is applied to a list
of secondary standards \citep{1992A&AS...96..379M,1994A&A...286..725H}
in order to compute their Gaussian full-width half-maximum (FWHM) on
the INT frames.  We then select a reference image for each of the CCDs
with an average FWHM closest to, but less than, 1.67 arcsec (5 WFC
pixels).  Any images exhibiting a poorer seeing than 5 pixels are
removed at this stage (accounting for $\sim 10$ images per CCD).
Images that show high ellipticity in their PSFs, further effects of
bad atmospheric conditions or instrumental/software deficiency are
also removed from the process.  Around $10-20$ images per CCD are
removed for these reasons.  The aligned images then have their PSFs
matched to that of the reference image using a PSF kernel size that
contains 90\% of the flux of objects in the reference image.  The
matching is carried out by applying {\tt psfmatch} to a
list of stars selected from the Magnier et al. catalogues which have good
PSFs and are resolved in all the images.

An estimate of the background (including the unresolved M31 light) is
produced for each PSF-matched image to enable background
subtraction. The background image is constructed by passing a
$49\times49$ pixel sliding filter, using IRAF's {\tt median} function.
Within the data there are a significant number of pixels that contain
counts arising from the extended diffraction structure of saturated
sources within, or close to, the observed fields. Image masks are
therefore constructed to identify these regions to the detection
pipeline.  A 24 pixel border around each CCD is excluded as the {\tt
median} function does not work reliably near the edges of images.  The
masks account for a further loss of 4.5\% of the data.  Table~\ref{lostpix} 
summarises the amount of data lost.

\begin{table}
\caption{Data lost due to the trimming of the aligned images and the
masking of regions of bad data.}
\label{lostpix}
\begin{tabular}{llll}
\hline
CCD     & \% lost by & \% lost by & Total \% \\
        & trimming   & masking    & lost \\
\hline
North 1 & 4.57\%     & 4.23\%     & 8.80\% \\
North 2 & 4.80\%     & 3.89\%     & 8.69\% \\
North 3 & 4.25\%     & 4.27\%     & 8.52\% \\
North 4 & 3.89\%     & 4.96\%     & 8.85\% \\
South 1 & 3.37\%     & 3.68\%     & 7.05\% \\
South 2 & 4.96\%     & 7.18\%     & 12.14\% \\
South 3 & 5.01\%     & 4.65\%     & 9.66\% \\
South 4 & 5.51\%     & 5.58\%     & 11.09\% \\
\hline
\end{tabular}
\end{table}

\section{Classical nova detection pipeline} \label{detectpipe}

\begin{figure*}
\includegraphics[width=405pt]{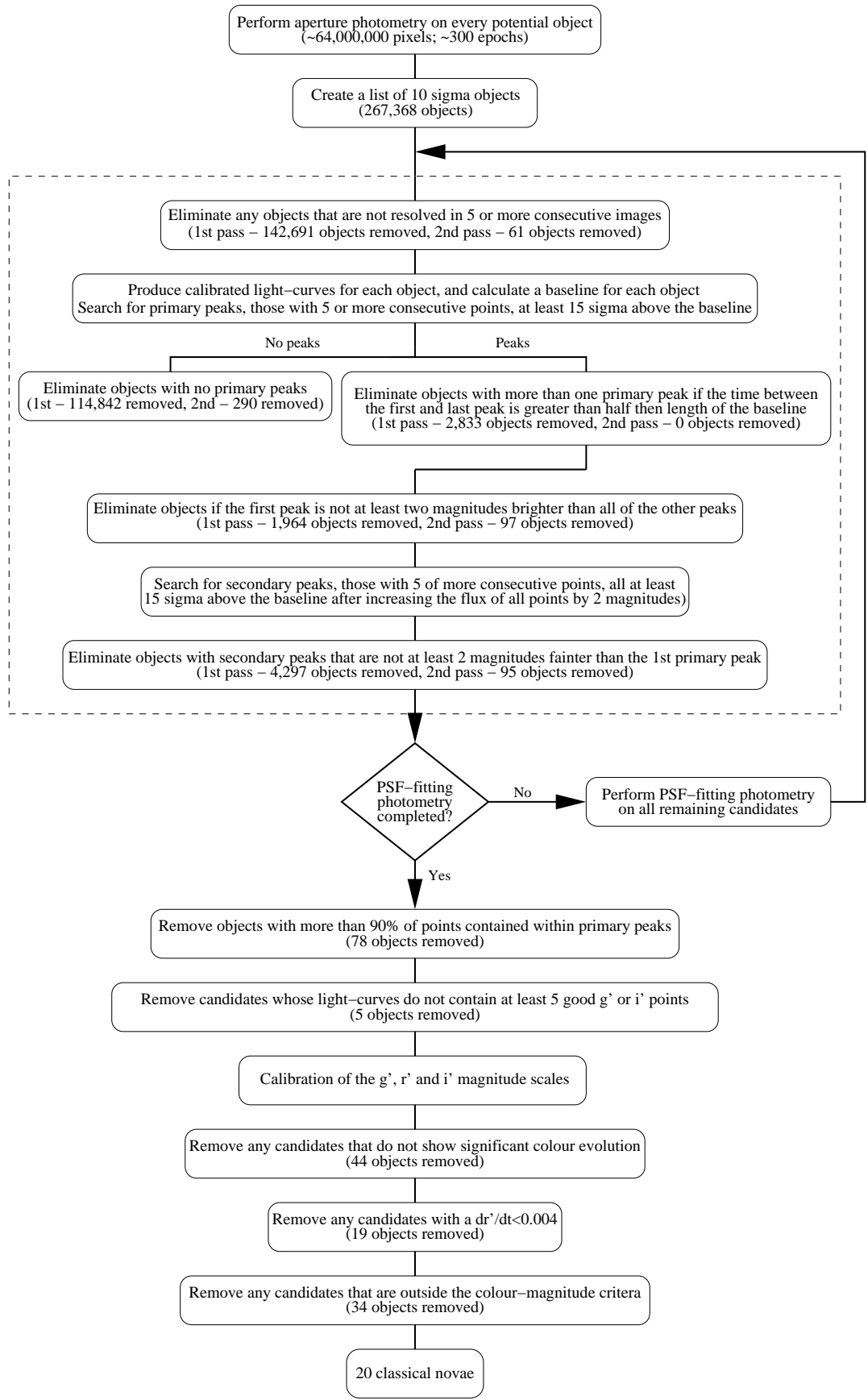}
\caption{A schematic diagram summarising our classical novae detection process.}
\label{pipeline}
\end{figure*}

After aligning and matching the $r'$ images, they are fed into our
nova detection pipeline.  The pipeline is written in C and C++ using
standard Unix/Linux libraries and the CFITSIO and CCFITS libraries.  A
flowchart of the pipeline, including the major steps and selection
criteria, as discussed in this section, is shown in
Figure~\ref{pipeline}.

\subsection{Standard star selection} \label{standards}

In order to detect objects which vary in flux, the first step in our
pipeline is to prepare a list of resolved standard stars which are
known not to vary in luminosity throughout the observations.  We later
calibrate the light-curves of our CN candidates relative to these
stars to eliminate any seeing effects in the data.  This selection is
carried out using the secondary standard stars from the Magnier et
al. BVRI catalogues of M31
\citep{1992A&AS...96..379M,1994A&A...286..725H}, which contain 485,425
objects.  As the catalogues make use of a different filter system from
that used by our survey, we must later use data provided by
the Cambridge Astronomical Survey Unit (CASU) INT Wide Field Survey
(WFS) to accurately calibrate our magnitude scales.  However, for the
purpose of relative calibration of the candidate light-curves, we
initially use a fiducial zero point of $r'=25$ to provide a rough
estimate of the magnitude scale of our data.

Candidates for non-varying stars are selected from the Magnier et al.
catalogues by virtue of their type (i.e. stars), reliability and
apparent magnitude.  Light-curves for each of the standard stars are
then produced using aperture photometry. Standard stars that contain
any points in their light-curves which correspond to saturated pixels
in any observation are immediately eliminated.  To define a sample of
standards which do not vary over the survey lifetime, we firstly make
the assumption that any variation in flux is due entirely to
extraneous factors such as seeing variations.  Then, for each epoch,
we calculate the mean flux correction such that the stars have
statistically the same flux as measured in the reference image.  We
then test the initial assumption that all the selected stars do not vary.  We
fit each standard star with a constant-flux light-curve.  If any of
the standard stars have fits with a reduced-$\chi^{2}>3$, then the star
exhibiting the poorest fit is assumed to have varied in flux.  This
process is iterated, removing the star with the poorest fit,
re-computing the statistical correction and refitting the light-curves
until all of the remaining stars have a reduced-$\chi^{2}\leq3$.

\subsection{Object definition}

To produce an initial list of CN candidates, we first create a list of
``objects'' for each observation epoch. We define an object to be a
resolved structure in a PSF-matched image with a flux at least
$10\sigma$ above the corresponding local median background flux. The
flux difference is designated to be the object flux.  Our object detection
routine, based upon the IRAF
{\tt daofind} package, allows us to deal with the strongly
varying background and uses median-filtered
images to estimate the local background.

We eliminate any objects from the candidate list if they do not have
$10\sigma$ detections for at least five consecutive observations.
This allows us to eliminate from our candidate list any rapidly
variable objects (which are unlikely to be novae), cosmic rays,
contamination from bad pixels and many of the effects of the extended
structure of unmasked saturated objects.  When comparing objects
between different images, some leeway is needed in the position of the
objects to allow for alignment and centring errors. We allow $\pm1$
pixel for the maximum error in our custom centring algorithm and $\pm
0.5$ pixels for the maximum error in the image alignment. In fact the
majority of the images are aligned to within $\pm0.1$ pixels across
the CCD.  Thus we treat objects on different frames which are
positioned to within $\pm2$ pixels as the same object.  Giving this
amount of leeway results in a small number of multiple detections
arising from nearby contaminating variable sources, particularly when
the contaminants are bright. However these duplications are easily
identified and removed at a later stage.

\subsection{Preliminary nova selection} \label{select}

We aim to define a set of selection criteria which are general in the
sense of not making unnecessary assumptions about CN light-curve
morphology, since we want to minimise the risk of biasing against the
detection of certain CN speed classes in favour of others. Our
starting point is the catalogues of Galactic nova light-curves
compiled by \citet{1981PASP...93..165D} and
\citet{1987A&AS...70..125V} and the ``ideal light-curve''
\citep{1960stat.book.....M}. We wish to derive a set of criteria
that allow us to select ``CN-like'' light-curves from our object list.
This is a difficult task since, for example, moderate speed class
novae, such as T Aurigae 1891 \citep{1941PASP...53..102M} and DQ
Herculis 1934 \citep{1936ApJ....84...14A}, exhibit minima $7-10$
magnitudes deep during the transition stage. There is also a clear
correlation between the morphology of CN light-curves and their rate
of decline.  These difficulties are further compounded for
extragalactic nova detection since it is possible that the transition
stage of a moderate speed class nova may be obscured by the host
galaxy's own surface brightness.  As a result, these novae may appear to
have multiple peaks.

Given the different behaviours of CN light-curves and the extended range of
light-curves due to potential contaminating objects, it is not possible, a
priori, to define objective selection criteria that are truely effective.  Thus,
the development of the algorithm used to isolate a sample of CN light-curves
involved an iterative process, in which the results of visual inspection of a
subset of light-curves surviving each stage in the selection-pipeline were used
to refine the selection criteria.  To ensure that the objective nature of the
selection criteria was preserved, no more than $10\%$ of the candidates 
surviving at each stage in the selection-pipeline were inspected visually.  It 
was also the case that the feedback from the visual inspection to the refinement 
of the selection criteria was concerned primarily with reducing the 
contamination of the CN candidate sample by various classes of variable star. 
Only at the final stages, where colour cuts are adopted, is it necessary to 
inspect all the candidates, though by this stage the ``obvious'' CN candidates 
are reasonably clearly delineated in magnitude and colour from the other
light-curves.

Our first task in the selection is to produce calibrated light-curves
for each of the remaining objects in our candidate list.  At this
preliminary stage the light-curves are produced using aperture
photometry and calibrated using the standard stars previously located.
For each light-curve we calculate a baseline flux by taking the minimum value
obtained from a sliding seven-epoch mean. We also require that the 7
consecutive points which define the baseline each lie within $3\sigma$ of the
baseline value. Windows of 7 points containing saturated data, or points lying
outside the $3\sigma$ limit, are discarded.  If any candidate's light-curve
contains no valid windows (i.e. it isn't possible to calculate a
baseline flux) then that candidate is discarded.

To characterise CN light-curves, we introduce the notion of primary and
secondary peaks. CNe can have complex light-curve structures but are
generically characterised by an initial large peak (the primary peak)
which, in some cases, may be followed by one or more lesser
(secondary) peaks. Secondary peaks tend to be at least 2 magnitudes
fainter than the primary.  The peaks themselves can exhibit some
sub-structure and therefore care is required in defining what constitutes a 
peak.

We define a primary peak as being bounded by points at least
$15\sigma$ above the baseline flux.  The other points within the
primary peak must either lie at least $15\sigma$ above the
baseline or within $3\sigma$ of the previous $15\sigma$ point
(in which case they are regarded as ``substructure'' points).  A
primary peak may contain any number of substructure points as long as
there are never more than 3 consecutive substructure points.  Finally
a primary peak must contain at least five points overall.  At this
stage we are able to discard the majority of candidates as they do not
contain any primary peaks.  We keep for the moment candidates which
contain one or more primary peaks.

For the surviving light-curves we calculate the characteristic width of
each  primary peak.  We define the end of each primary peak as the
first point following the peak maximum at which the flux of the object
drops below $3\sigma$ above the baseline, or the final observation if
this occurs first.  Using a similar definition for the start of each
primary peak, we are able to specify the size of each peak. Overlapping
primary peaks are then re-assigned as a single primary peak.  At
this stage, for light-curves with more than one primary peak we
require that the time between the maximum flux of the first and last
primary peak be less than half the total baseline time of the survey.
This criterion is introduced in order to eliminate ``contained''
periodic variables.

From our studies of a large proportion of all past Galactic nova
light-curves, we have noticed that maxima occurring after the
transition phase of the nova are always at least two magnitudes
fainter than its maximum light.  This presents a simple way to
eliminate the majority of the multiple primary peak candidates: we
require that all primary peaks after the initial one be at least 2
magnitudes below the first peak.

From the remaining light-curves we search for secondary peaks.  We
define a secondary peak as being 5 or more consecutive points lying at
least $15\sigma$ above the baseline {\em after increasing the flux of
each data point by two magnitudes}\/. Points already associated with
primary peaks are excluded from the search.  We then eliminate
candidates unless their secondary peaks are at least 2 magnitudes
fainter than the first primary peak.

\subsection{Relative PSF-fitted photometry}

The result of the CNe detection pipeline at this stage is a
preliminary catalogue of 741 candidates, compared to the 9,835
strongly variable objects (at least one primary peak) originally
identified from 267,368 $10\sigma$ objects detected within the two
fields. The contaminants within this preliminary catalogue are
expected to be mainly long-period variables such as Miras. Further
cuts require accurate, multi-colour photometry.  This is performed for
each surviving candidate in all three colours ($g'$, $r'$ and $i'$)
for all three seasons.

Accurate photometric measurements are obtained by PSF fitting rather
than relying on the aperture photometry used until now. The PSF
fitting handles the strongly variable background much more robustly
than the simple aperture photometry technique.  This is carried out
for each of the selected Magnier et al. standards, as well as for the
CN candidates.  Instrumental magnitudes are extracted using the {\tt
psf} and {\tt peak} routines within IRAF's {\tt daophot} package. After this
more accurate calibration, the flux
stability of the standards is re-checked using the procedure outlined
in Section~\ref{standards}.  In total we identify 650 standards with stable  
$r'$ light-curves, the number of standards per CCD ranges
from 9 (South-field CCD2) to 202 (South-field CCD3).

\begin{table*}
\begin{minipage}{170mm}
\caption{The effect of each stage of our selection pipeline upon the
classical nova catalogue.  These steps are described in
Section~\ref{detectpipe}.}
\label{elim}
\begin{tabular}{llllllllll}
\hline
Pipeline                   & \multicolumn{4}{c}{North-field}       &
\multicolumn{4}{c}{South-Field}       & All \\stage                      & CCD1  
  & CCD2    & CCD3    & CCD4    & CCD1    & CCD2    & CCD3    & CCD4    & CCDs 
\\ \hline
Pixels                     & $8\times10^{6}$ & $8\times10^{6}$&
$8\times10^{6}$& $8\times10^{6}$& $8\times10^{6}$& $8\times10^{6}$&
$8\times10^{6}$& $8\times10^{6}$ & $6.4\times10^{7}$\\$10\sigma$ objects         
& 29,423  & 29,931  & 39,921  & 33,125  & 24,167  & 28,220  & 37,086  & 45,495  
& 267,368\\
                         & \multicolumn{9}{c}{{\em Pipeline $1st$ 
pass -- aperture photometry}}\\
5 consecutive detections   & 17,976  & 15,901  & 
19,292  & 18,292  & 12,006  & 16,258  & 22,433  & 20,519  & 124,677\\
$\geq1$ primary peak       & 1,366   & 898     & 1,073   & 1,215   & 703     & 
821     &2,146   & 1,613   & 9,835\\
Periodicity test           & 1,036   & 686     & 731  
   & 908     & 468     & 571     & 1,524   & 1,078   & 7,002\\
Primary peak height        & 794     & 539     & 464     & 659     & 318     
& 415     & 1,126   & 723     & 5,038\\
Secondary peak height      & 145     & 77      & 62   
   & 122     & 37      & 66      & 142     & 90      & 741\\                     
      & \multicolumn{9}{c}{{\em Pipeline $2nd$ pass -- PSF-fitting 
photometry}}\\
5 consecutive detections   & 135     & 66      & 59      & 110     
& 35      & 62      & 132     & 81      & 680\\
$\geq1$ primary peak       & 66   
   & 43      & 38      & 59      & 29      & 44      & 61      & 50      &
390\\
Periodicity test           & 66      & 43      & 38      & 59      & 29     
 & 44      & 61      & 50      & 390\\
Primary peak height        & 59      & 33  
    & 26      & 49      & 16      & 24      & 52      & 34      & 293\\
Secondary peak height      & 35      & 21      & 21      & 35      & 12      & 
16      & 37      & 21      & 198\\                           & 
\multicolumn{9}{c}{{\em Further candidate elimination stages}}\\
$<90\%$ of points in peaks & 22      & 7       & 7       & 21      & 8       & 
13      & 29      & 13      & 120\\
5 $g'$ or $i'$ points      & 22      & 5      
 & 7       & 20      & 8       & 12      & 28      & 13      & 115\\
Colour evolution           & 12      & 4       & 7       & 8       & 5       & 8 
      & 17      & 10      & 71\\
Rate of decline            & 9       & 4       & 
5       & 5       & 3       & 5       & 14      & 7       & 
52\\
Colour--magnitude criteria & 4       & 4       & 1       & 3       & 0       
& 0       & 6       & 2       & 20\\\\
{\bf Final candidates}     & {\bf 4} & 
{\bf 4} & {\bf 1} & {\bf 3} & {\bf 0} & {\bf 0} & {\bf 6} & {\bf 2} & {\bf 
20}\\
\hline
\end{tabular}
\end{minipage}
\end{table*}

We can now re-apply the
criteria previously used in Section~\ref{select} to eliminate further
candidates, taking advantage of the more accurate photometry. This re-processing
proves very important in order to remove spurious candidates allowed by
aperture photometry. A further 543 candidates are eliminated -- more than 70\%
of the surviving sample.  A breakdown of the candidates eliminated at each stage
in the pipeline is given in Table~\ref{elim}. The higher photometric
reliability allows us to introduce further selection criteria.  We require that
the light-curves should not have more than 90\% of their points in primary
peaks. This is essentially a second ``containment'' criterion which allows us to
eliminate long period variables.   To investigate colour
evolution we further require that the candidates
comprise at least 5 points in either $g'$ or $i'$.

\subsection{Photometric calibration and colour selections} \label{coltrans}

To calibrate our instrumental $r'$, $i'$ and $g'$ magnitudes we used
the calibrated zero points for the INT WFC calculated by the CASU WFS
team.  Their zero points, calculated for many of the nights on which
POINT-AGAPE observations were also taken, allow us to test our
relative photometry.

Since the colour of CNe is known to vary strongly throughout
the eruption (especially just after maximum
light), we demand that the $r'-i'$ and $g'-r'$ colours of our
candidates show significant colour evolution. Accordingly we compute
the reduced-$\chi^2$ for a constant colour fit and reject those
light-curves where the reduced-$\chi^2< 3$. This cut essentially rejects
light-curves unless they exhibit colour evolution at a high level of
significance.

\begin{figure}
\rotatebox{270}{\includegraphics[width=240pt]{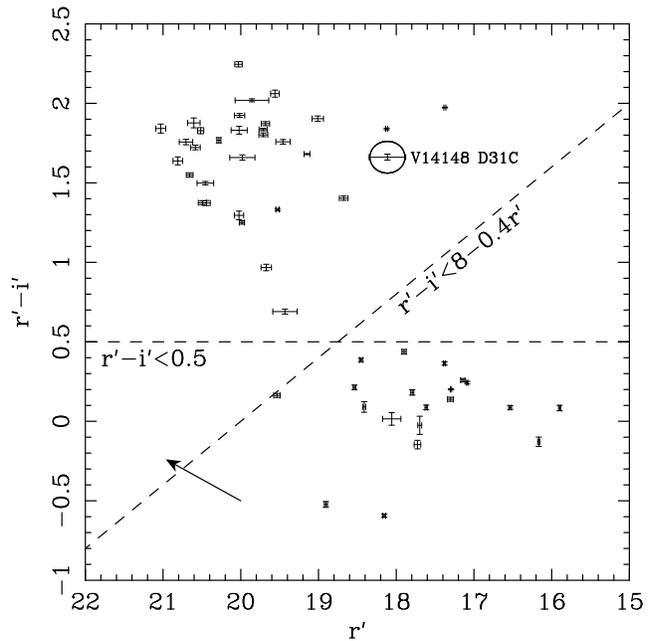}}
\caption{A colour--magnitude diagram showing the 52 candidates
remaining prior to colour selection.  The 20 objects located within
the region defined by Equations~(\ref{colmag1}) and (\ref{colmag2}) are
the classical novae discovered by this survey.  The group of objects
centred at $r'\sim20, r'-i'\sim1.5$ appear to be mainly Mira
variables.  A notable non-nova candidate is indicated, this is
discussed in more detail in Section~\ref{borderl}.  The arrow
indicates the direction of the reddening vector.  The apparent
discrepancy in the relative sizes of the $r'$ and $r'-i'$ errors seen
on some of the points, e.g. V9205~D31C can be understood by
considering that the $r'$ error is drawn solely from the brightest
observation (which may have been on a night where no $i'$ data were
taken or when the object was not visible in $i'$) where as the $r'-i'$
points, and hence their errors, are drawn from the average colour
following maximum light.}
\label{colmag}
\end{figure}

To eliminate candidates from the catalogue which decay too slowly to
be CN, we introduce a rate-of-decline criterion.  Using the speed
class definitions in Table~\ref{speed} as a guide, and applying them
to the $r'$ band, we remove light-curves where $dr'/dt<0.004$~mag~day$^{-1}$
(equivalent to $t_2 >500$ days).  At this stage we are left with
the 52 candidate light-curves which are plotted on the colour-magnitude diagram 
in Figure~\ref{colmag} which shows $r'$ at peak versus $r'-i'$ near peak. 
Specifically, $r'-i'$ in Figure~\ref{colmag} represents the mean colour of the 
object in the magnitude range $r'_{\max}<r'<r'_{\max}+3$.  The mean is used in 
order to average over colour variations near maximum light. In some cases, where 
the peak flux occurs during poor observing conditions, the mean $r'-i'$ is 
better determined than $r'$ at peak. The candidates are, for the most 
part segregated into two clumps, one at $r'\sim 20$, $r'-i' \sim 1.5$ and the 
other at $r'\sim 18$, $r'-i' \sim 0$. If the remaining candidates lie within 
M31, we expect them to suffer to varying degrees from extinction within that 
galaxy. However, as the $r'$ and $i'$ filters are relatively closely spaced in 
wavelength, we expect that the change in $r'-i'$ due to extinction is small,
compared to the decrease in both $r'$ and $i'$, and much smaller than
the change in $g'-r'$.  Therefore, the use of colour-magnitude criteria
to eliminate further CN candidates is practical. In order to
eliminate bogus candidates from the list, we introduce two further
criteria:
\begin{equation}
r'-i'<0.5
\label{colmag1}
\end{equation}
\begin{equation}
r'-i'<8-0.4r'
\label{colmag2}
\end{equation}
The rationale behind
Equation~(\ref{colmag1}) is that we expect novae to have approximately
equal brightness in all bands at around maximum, after which they
become bluer as the eruption develops. This cut effectively eliminates
the clump at $r'\sim 20$, $r'-i' \sim 1.5$, which appears mostly to
comprise Mira variables. The second colour criterion in
Equation~(\ref{colmag2}) is almost orthogonal to the line joining
the two clumps and also to the expected direction of the reddening
vector indicated in Figure~\ref{colmag}. This cut ensures that, whilst
we may potentially miss CNe due to reddening, extinction should not
cause Miras or similar objects to be mistaken for CNe. This second cut
also deals with novae whose maximum light may have been missed; we
would expect these novae to appear fainter and bluer.  A summary of
all of the selection criteria, and their impact for the number of
surviving candidates at each stage for each CCD is given in
Table~\ref{elim}.

\subsection{Astrometry}

To calculate the astrometric positions of each CN, we first
compute the pixel positions of each nova whilst at or near maximum
light using the apphot {\tt center} function.  The pixel coordinates
of the selected Magnier et al. standards are determined in the
reference frames using the {\tt wcsctran} package.  Image position
solutions for each CCD are obtained using matched celestial and pixel
coordinate lists for the novae within the {\tt ccmap} package to
calculate second-order geometric transformations.  Finally, the
celestial coordinates for each CN are calculated using the computed
solutions with the {\tt cctran} package.

\section{The catalogue} \label{catalogue}

\begin{table*}
\caption{}
\vbox to220mm{\vfil Landscape table to go here.
\vfil}
\label{canddata}
\end{table*}

Following the implementation of our nova detection pipeline and all
the candidate selection criteria, we have identified 20 CN candidates.
The positions of each of the CNe and further information 
is tabulated in Table~\ref{canddata}.  The $dr'/dt$ parameter is
estimated from the general slope of the $r'$-band light-curve between
the brightest observation and the observation closest to 2 magnitudes
fainter than the maximum.  The speed class of each CN is then
estimated using the definition given in Table~\ref{speed}.  However,
note that the various speed classes in Table~\ref{speed} are defined
for $V$-band light-curves, whilst we are applying them to $r'$-band
data.  Since novae become bluer as they decline, we expect a slight
over-estimation of the speeds of the CN relative to the $V$-band
definitions.  Table~\ref{SCDist} shows the distribution of our sample
of CNe with speed class.

\begin{table}
\caption{M31 classical novae speed class distribution.  See
Table~\protect\ref{speed} for the speed class definitions.}
\label{SCDist}
\begin{tabular}{lll}
\hline
Speed           & Number & CN observed before\\
class           & of CN  & maximum-light\\
\hline
Very Fast       & 1      & 0\\
Fast            & 3      & 2\\
Moderately Fast & 11     & 5\\
Slow            & 3      & 2\\
Very Slow       & 2      & 0\\
\hline
\end{tabular}
\end{table}

\begin{figure*}
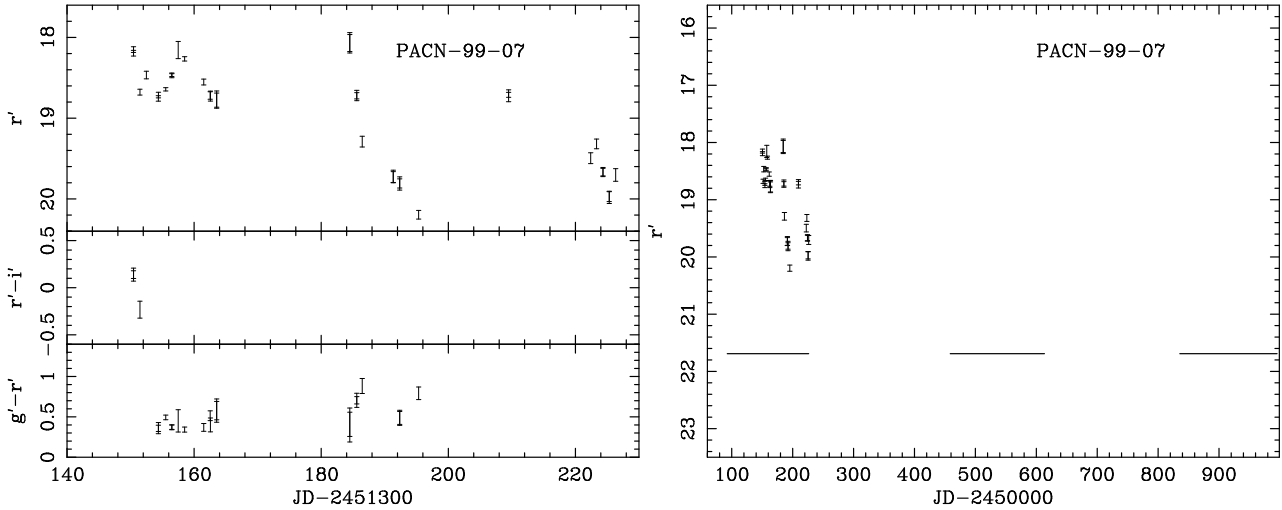

\rotatebox{270}{\includegraphics[width=190pt]{PACN-99-07_1.eps}}
\rotatebox{270}{\includegraphics[width=190pt]{PACN-99-07_2.eps}}
\caption{A ``very fast'' nova discovered in M31 from the POINT--AGAPE
survey.  The left-hand pane shows the $r'$-band light-curve around the
peak of the nova along with the $r'-i'$ and $g'-r'$ colours, where
available. The right-hand pane shows the complete $r'$-band
light-curve over the entire observation period, the horizontal lines
indicate the span of our observing seasons and the approximate
$r'$-band PSF-fitting magnitude sensitivity in the locale of the
CN. The automated classification of this CN as very fast arises from
the decay rate of the most prominent peak in $r'$, occurring around
JD$\, \simeq 2451485$~days. Inspection of the full activity around
peak indicates that the lightcurve better resembles a moderately fast
CN.}
\label{vfast}
\end{figure*}

\begin{figure*}
\rotatebox{270}{\includegraphics[width=190pt]{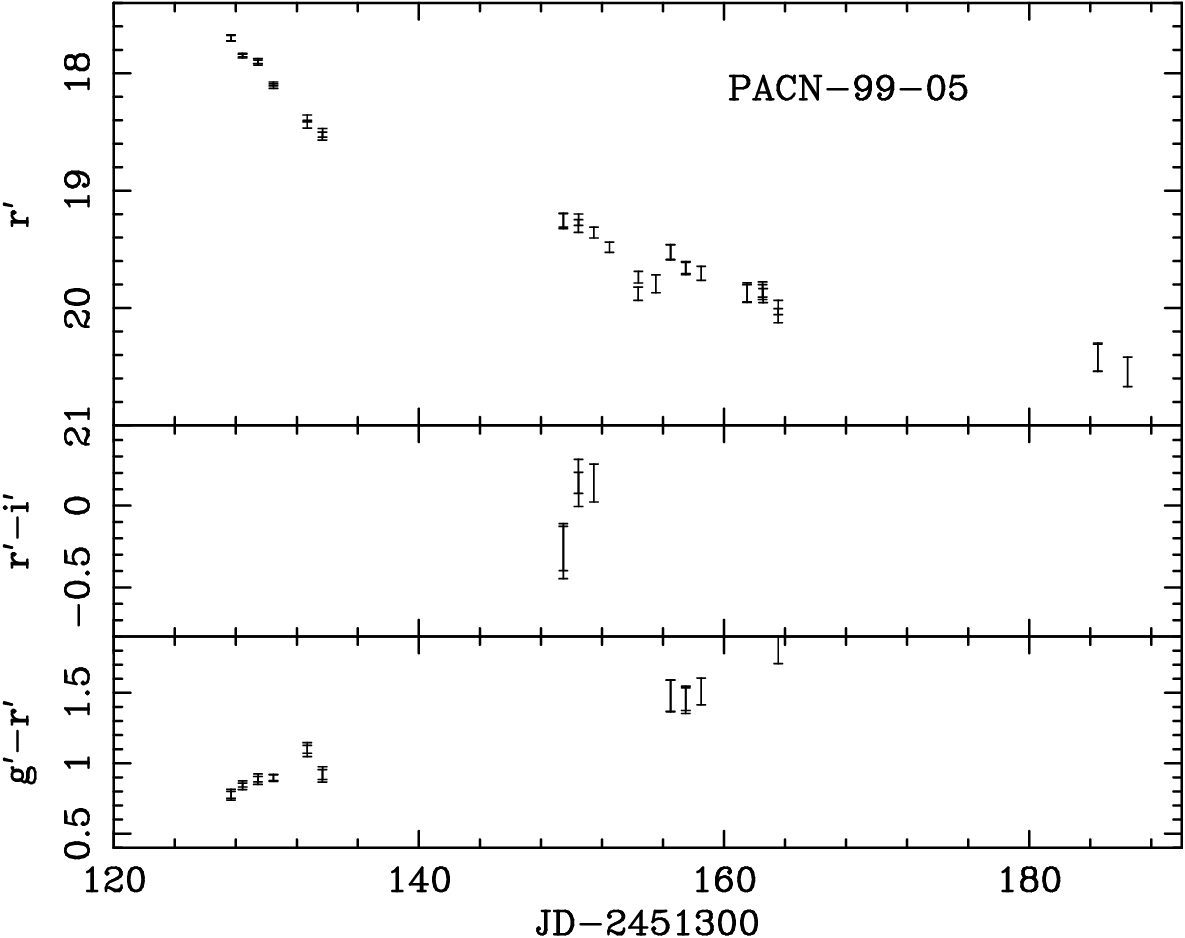}}
\rotatebox{270}{\includegraphics[width=190pt]{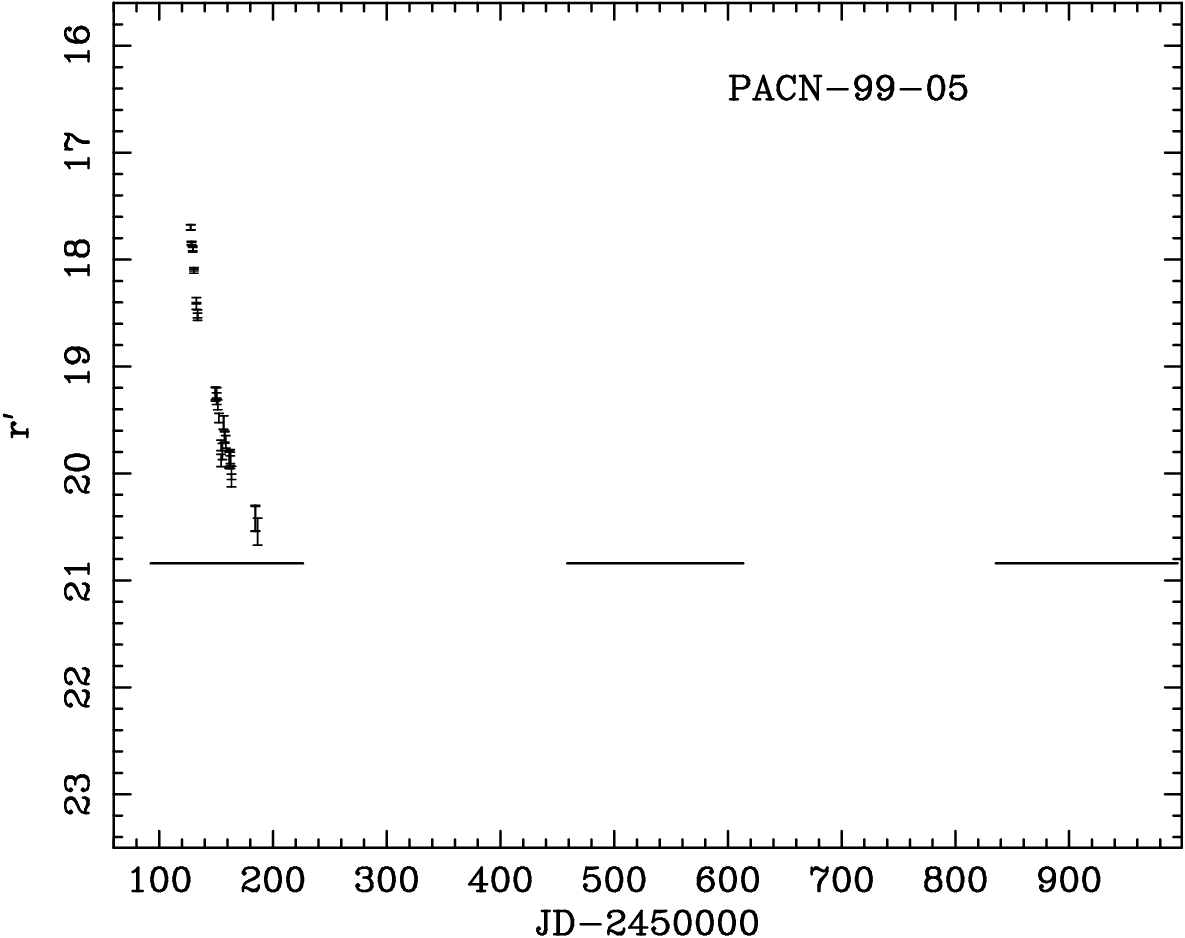}}
\rotatebox{270}{\includegraphics[width=190pt]{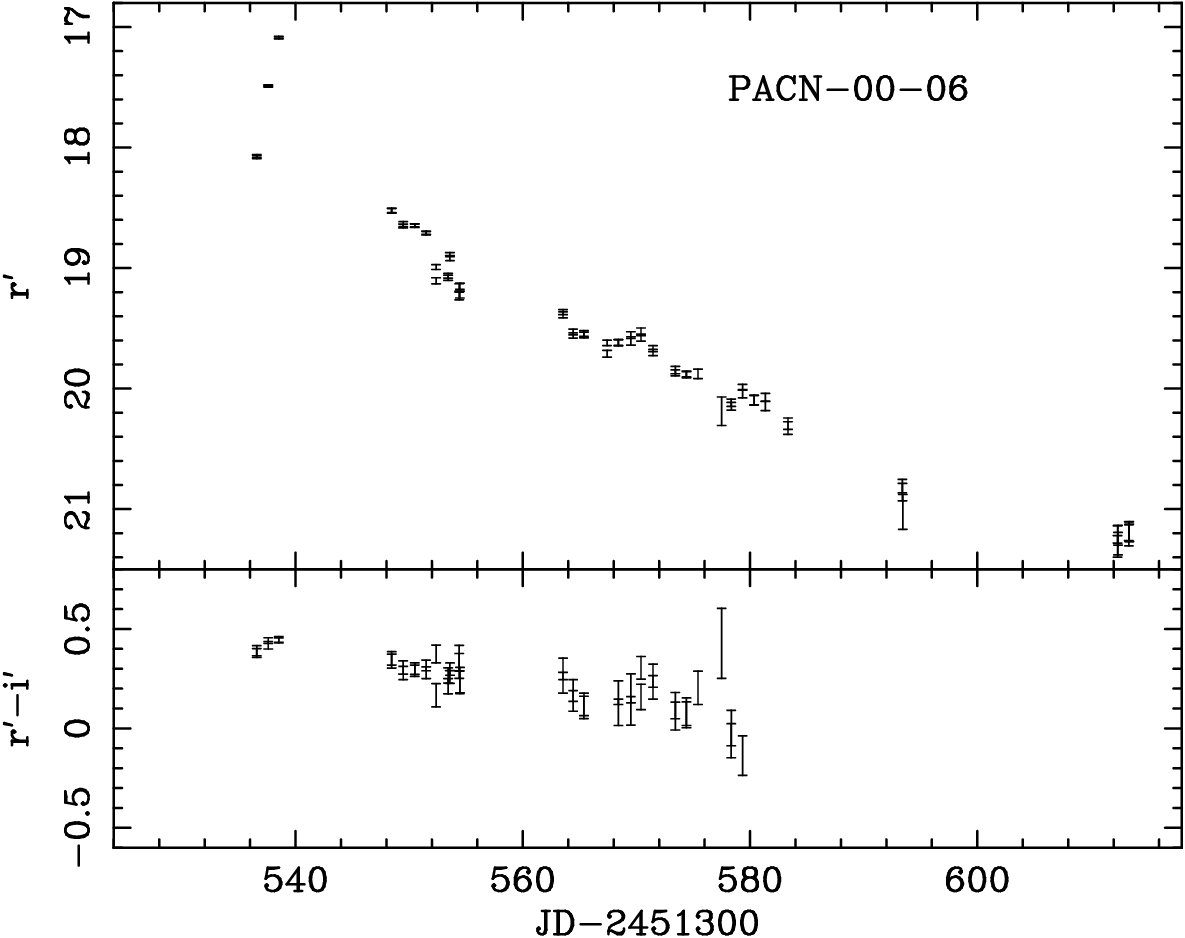}}
\rotatebox{270}{\includegraphics[width=190pt]{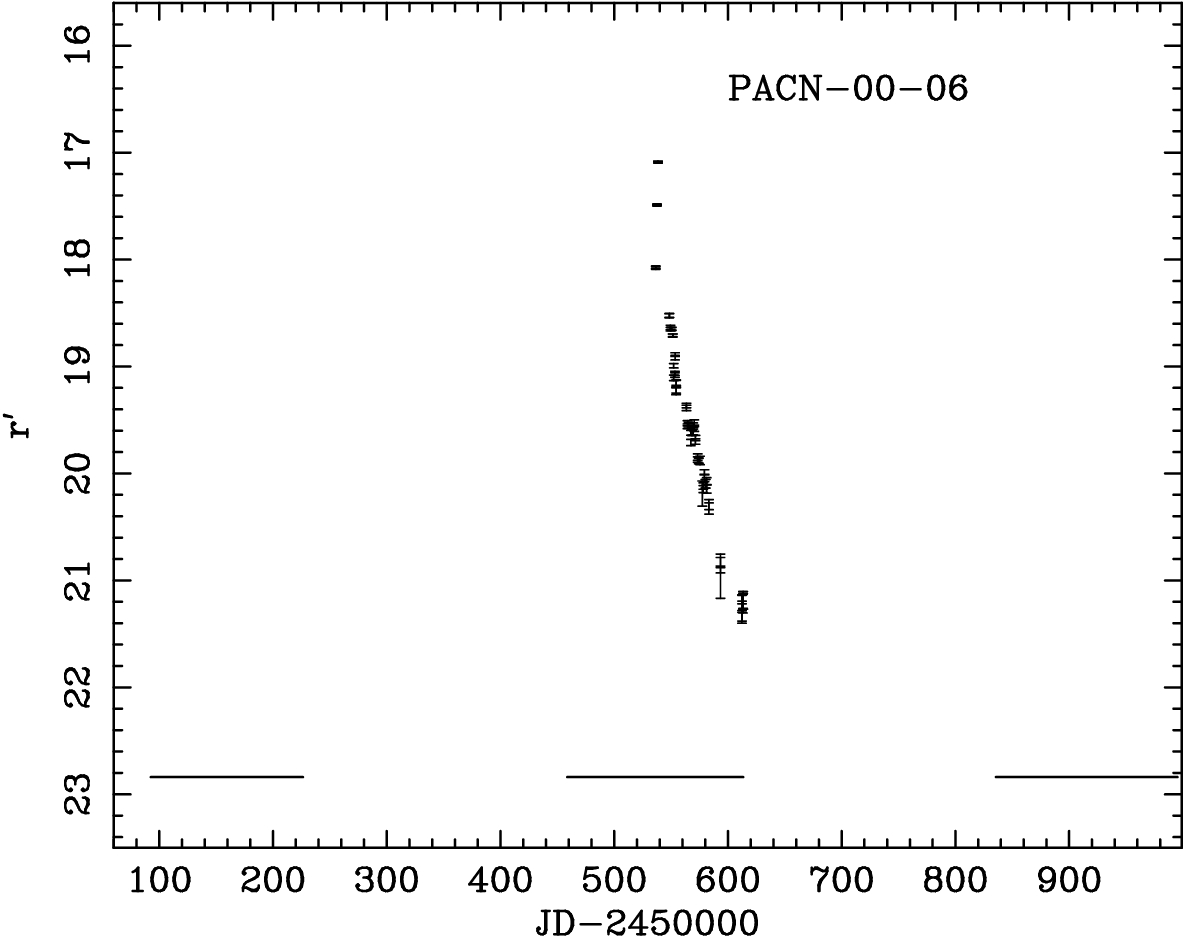}}
\rotatebox{270}{\includegraphics[width=190pt]{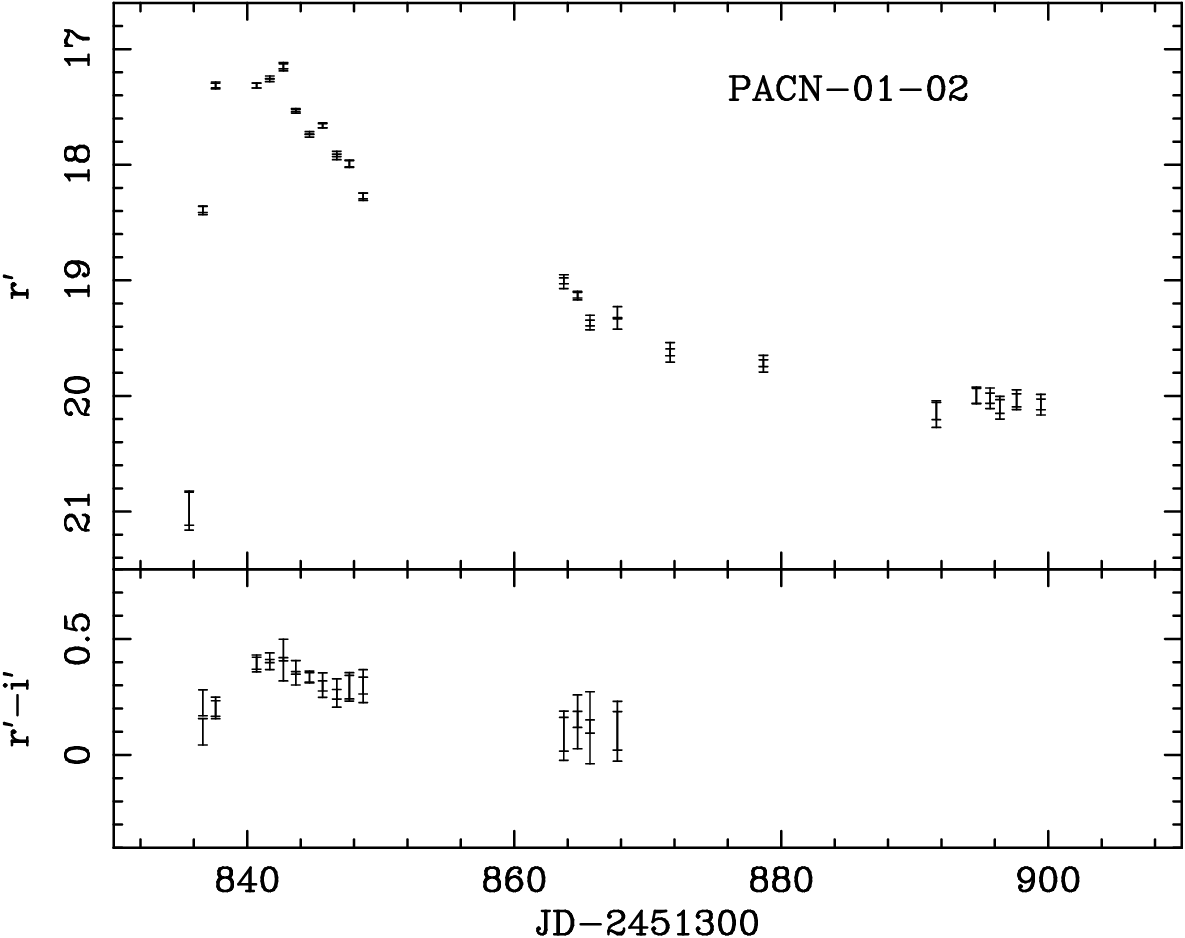}}
\rotatebox{270}{\includegraphics[width=190pt]{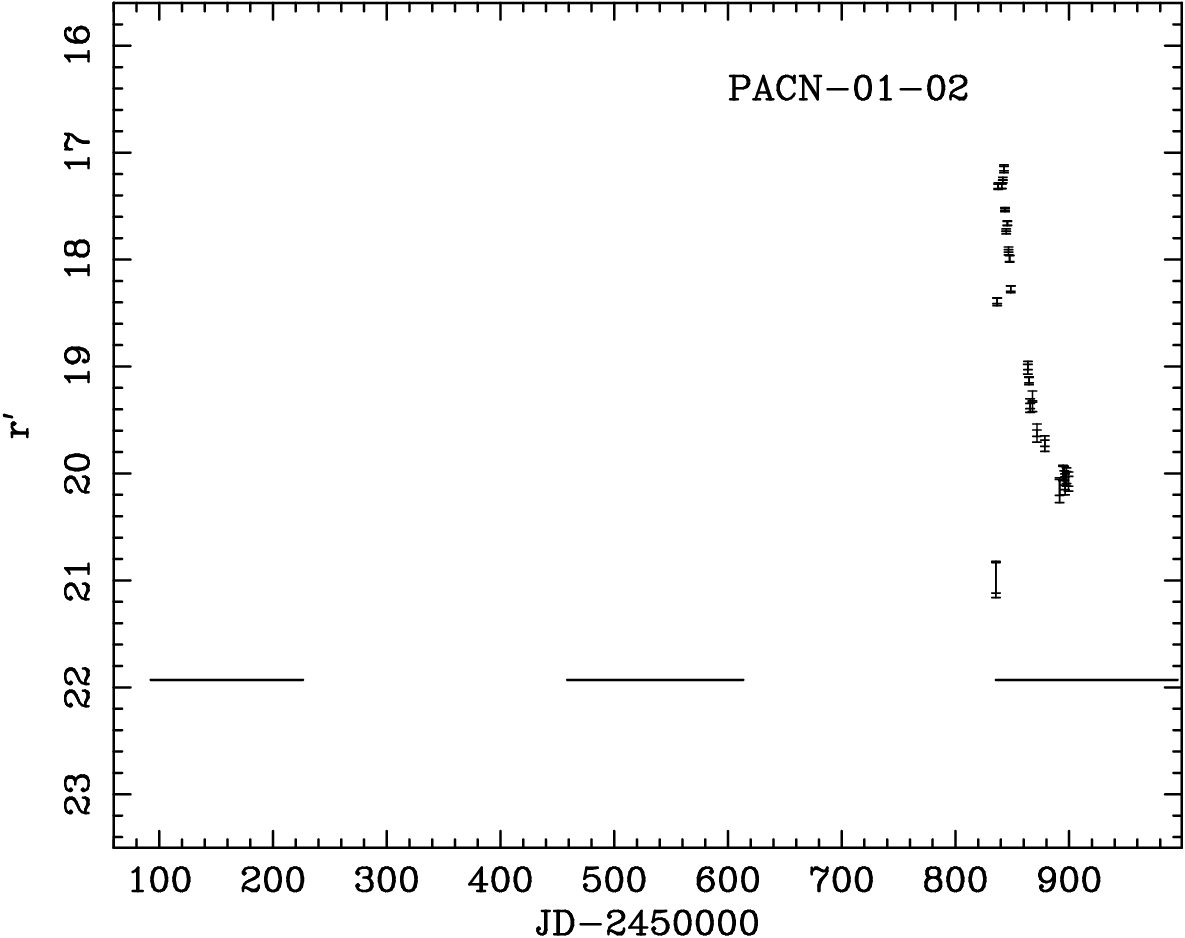}}
\caption{Fast novae in the POINT--AGAPE dataset. The panels are as for Figure~\ref{vfast}.}
\label{fast}
\end{figure*}

Figures~\ref{vfast} to \ref{vslow} provide the $r'$-band light-curves for each 
of the 20 CNe discovered and identified in Table~\ref{canddata}.  Also
shown are the $g'-r'$ and $r'-i'$ colours, where available. The span of our
 three observing seasons, and the approximate $r'$-band magnitude limit of
the PSF fitting are indicated by the horizontal lines in the $r'$-band
panels. The magnitude limit is determined in the immediate region of each
candidate by adding successively brighter artificial PSFs to the data until
they are recognised by the PSF-fitting routine. This is performed only on
our reference CCD frames (those with a seeing scale closest to 5 pixels), and so
represents only an approximate limit for data at other epochs. Where
points in the $r'-i'$ light-curves are apparently ``missing'' this is
usually due to the object being particularly blue. This, coupled with
the higher ($\sim1$ magnitude) zero-point of the INT CCDs in $i'$,
means that very blue objects such as novae are often unresolved in
$i'$-band observations.

The following subsections describe in some detail features of the
light-curves of selected CNe from each speed class.


\subsection{Very fast novae}
A lone ``very fast'' nova was identified by the pipeline, PACN-99-07,
which is plotted in Figure~\ref{vfast}. The most prominent peak in the
$r'$-band lightcurve of this CN (occurring around JD$\, \simeq
2451485$~days) exhibits a $dr'/dt=0.2$ equivalent to a $t_{2}\sim 20$
days. However, this peak appears to be substructure within a broader
outburst which seems to have skewed the estimation of the speed
class.  PACN-99-07 reaches a maximum magnitude of $r'=18.1\pm0.1$ -
much lower than would be expected for a very fast nova at, or near,
peak brightness at the distance of M31 (assuming no significant
extinction). This, together with the full behaviour of the lightcurve
around outburst, leads us to the conclusion that PACN-99-07 is in fact
a moderately fast speed class CN.

\subsection{Fast novae}
As shown in Table~\ref{SCDist} and Figure~\ref{fast}, three of the CNe 
discovered are fast novae, taking between $11-25$ days to decrease in brightness 
by two magnitudes from maximum light. Two of these fast CN (PACN-00-06 and 
PACN-01-02) have been caught in their final rise phase before maximum-light.
PACN-99-05 appears to have been first observed at or around maximum.

PACN-00-06 was observed four times during its final rise phase.  It
was first observed at on 19th October 2000 with $r'=18.08\pm0.01$,
before rising to $r'=17.09\pm0.01$ on 21st October.  The nova was
again observed two weeks later about 1.5 magnitudes
fainter. PACN-00-06 was followed for about 4 magnitudes.  PACN-00-06
was also observed by the Naini~Tal M31 microlensing survey group
\citep{2004A&A...415..471J}, and was designated by them CN~NMS-1.

PACN-01-02 was observed several times before maximum light, first on
15th August 2001. It brightened by 1.3 magnitudes until it reached a
maximum-light of $r'=17.14\pm0.03$ on 21st August.  The light-curve
was well sampled through maximum-light and into the initial decline
phase and was followed for around three magnitudes.

\subsection{Moderately fast novae}

\begin{figure*}
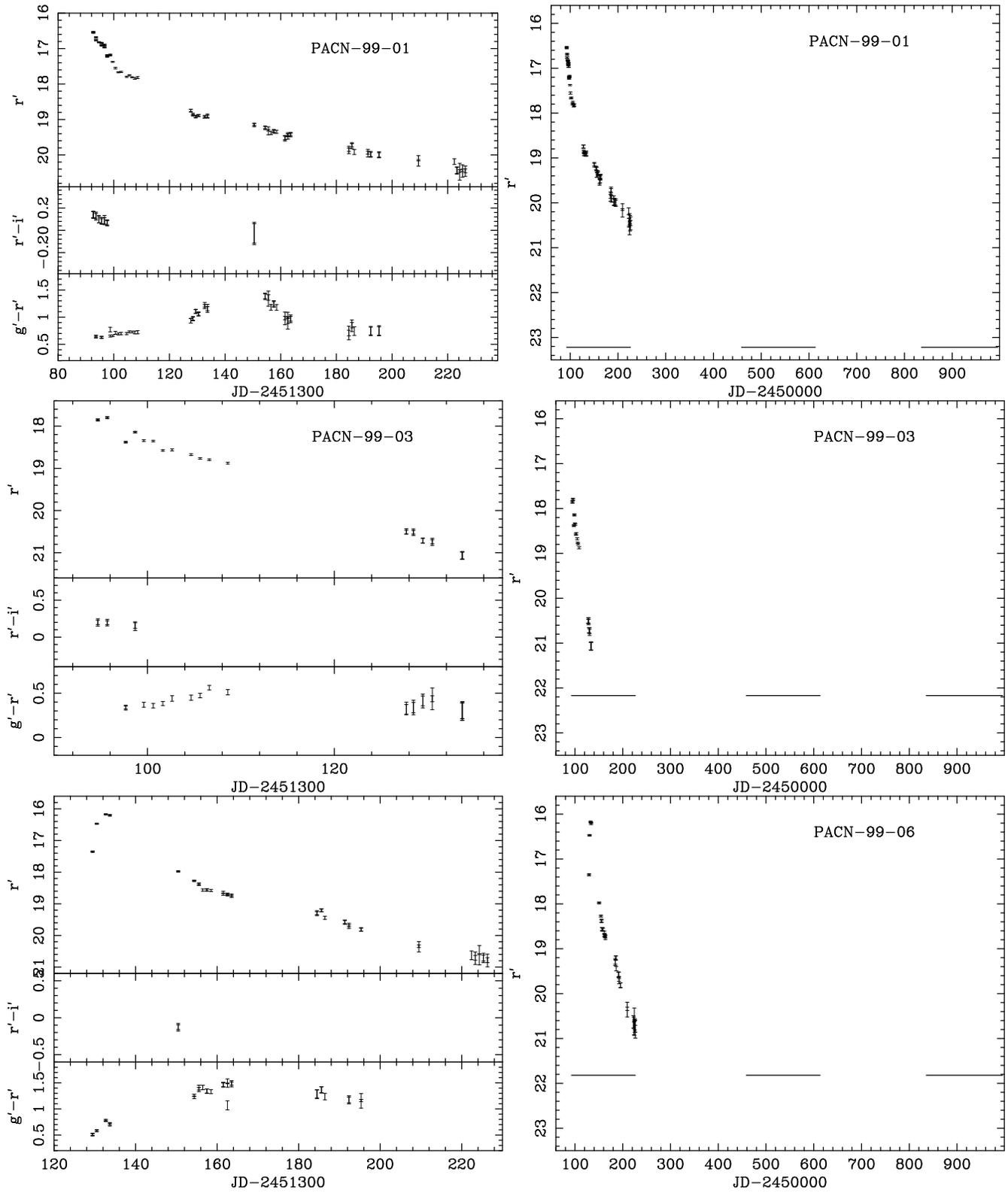

\rotatebox{270}{\includegraphics[width=198pt]{PACN-99-01_1.eps}}
\rotatebox{270}{\includegraphics[width=198pt]{PACN-99-01_2.eps}}
\rotatebox{270}{\includegraphics[width=198pt]{PACN-99-03_1.eps}}
\rotatebox{270}{\includegraphics[width=198pt]{PACN-99-03_2.eps}}
\rotatebox{270}{\includegraphics[width=198pt]{PACN-99-06_1.eps}}
\rotatebox{270}{\includegraphics[width=198pt]{PACN-99-06_2.eps}}
\caption{Moderately fast novae in the POINT--AGAPE dataset. The panels are as for Figure~\ref{vfast}.}
\label{mfast}
\end{figure*}
\addtocounter{figure}{-1}
\begin{figure*}
\rotatebox{270}{\includegraphics[width=198pt]{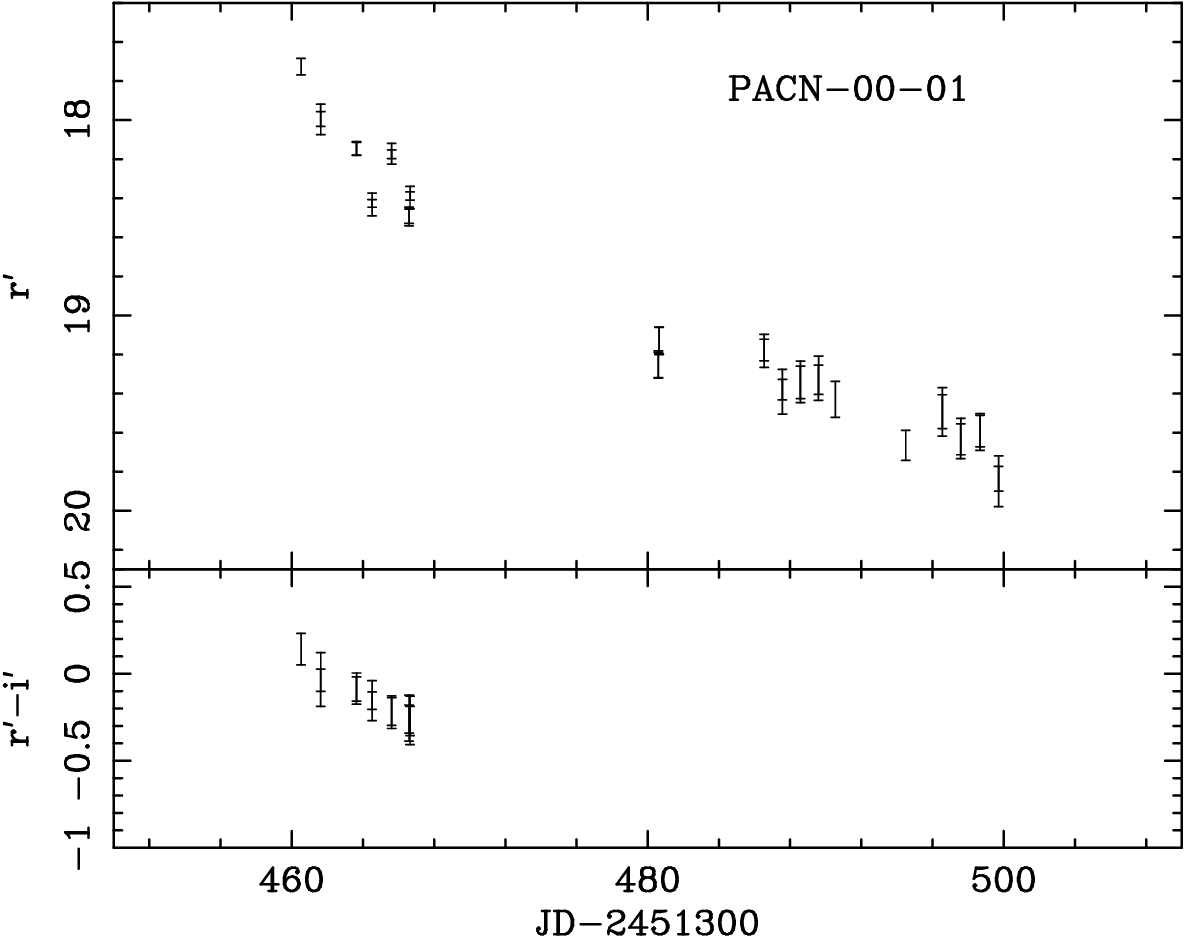}}
\rotatebox{270}{\includegraphics[width=198pt]{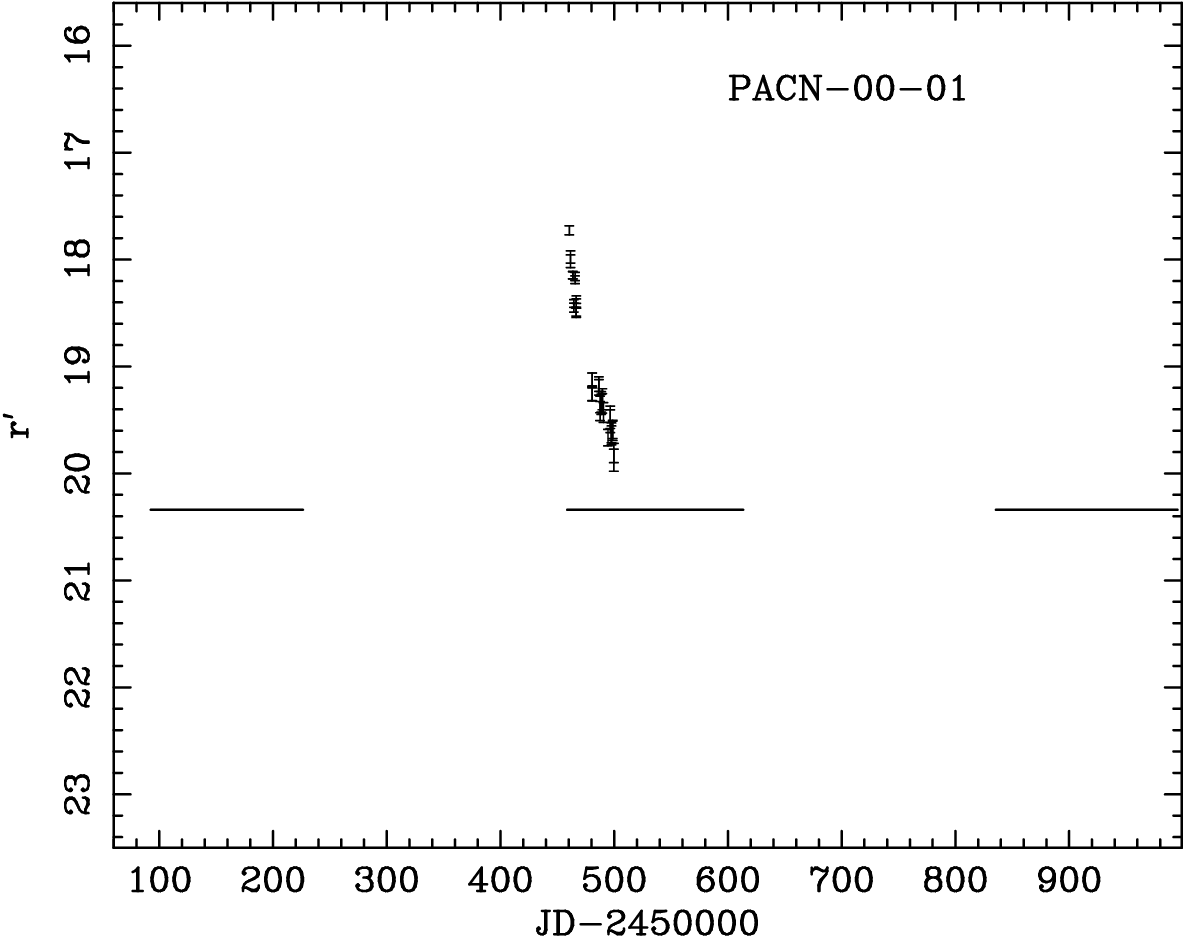}}
\rotatebox{270}{\includegraphics[width=198pt]{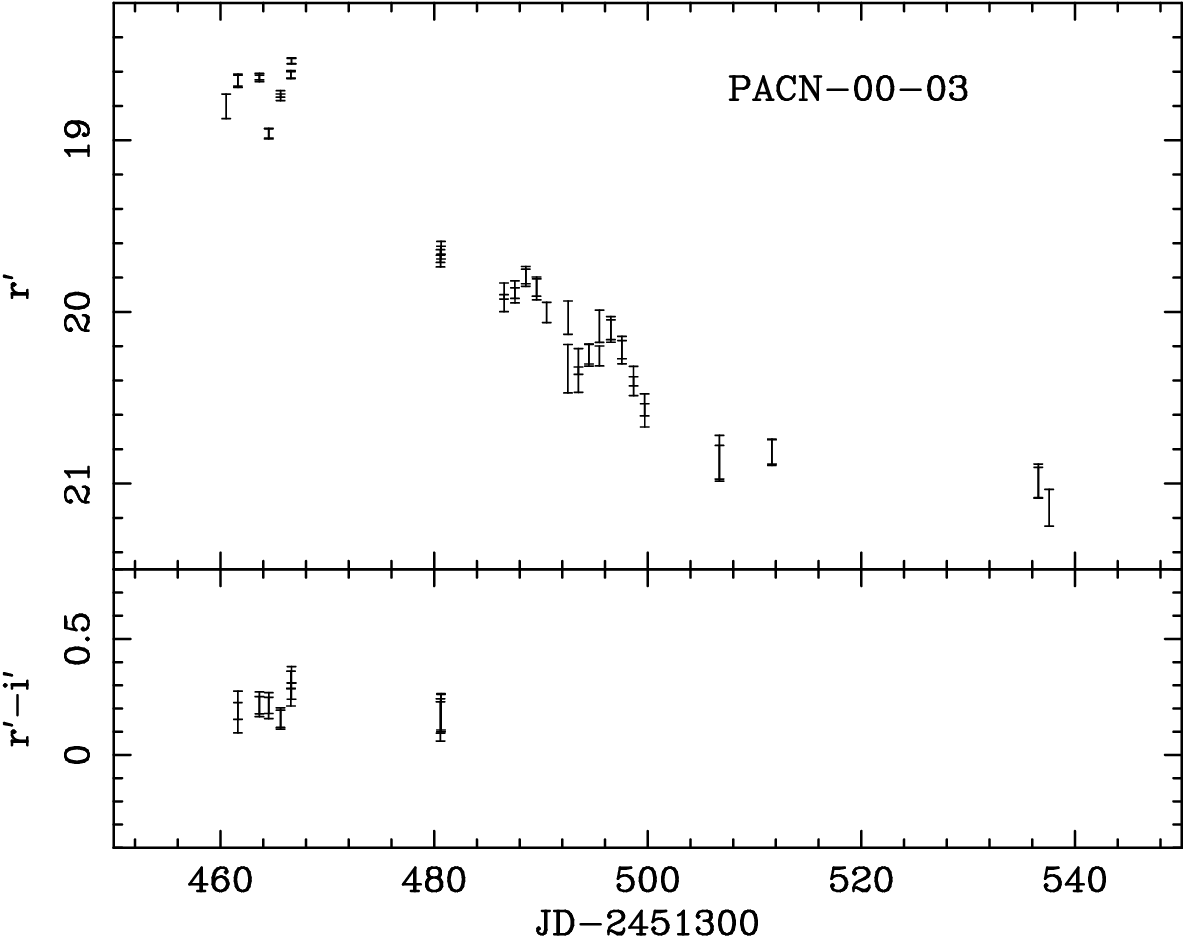}}
\rotatebox{270}{\includegraphics[width=198pt]{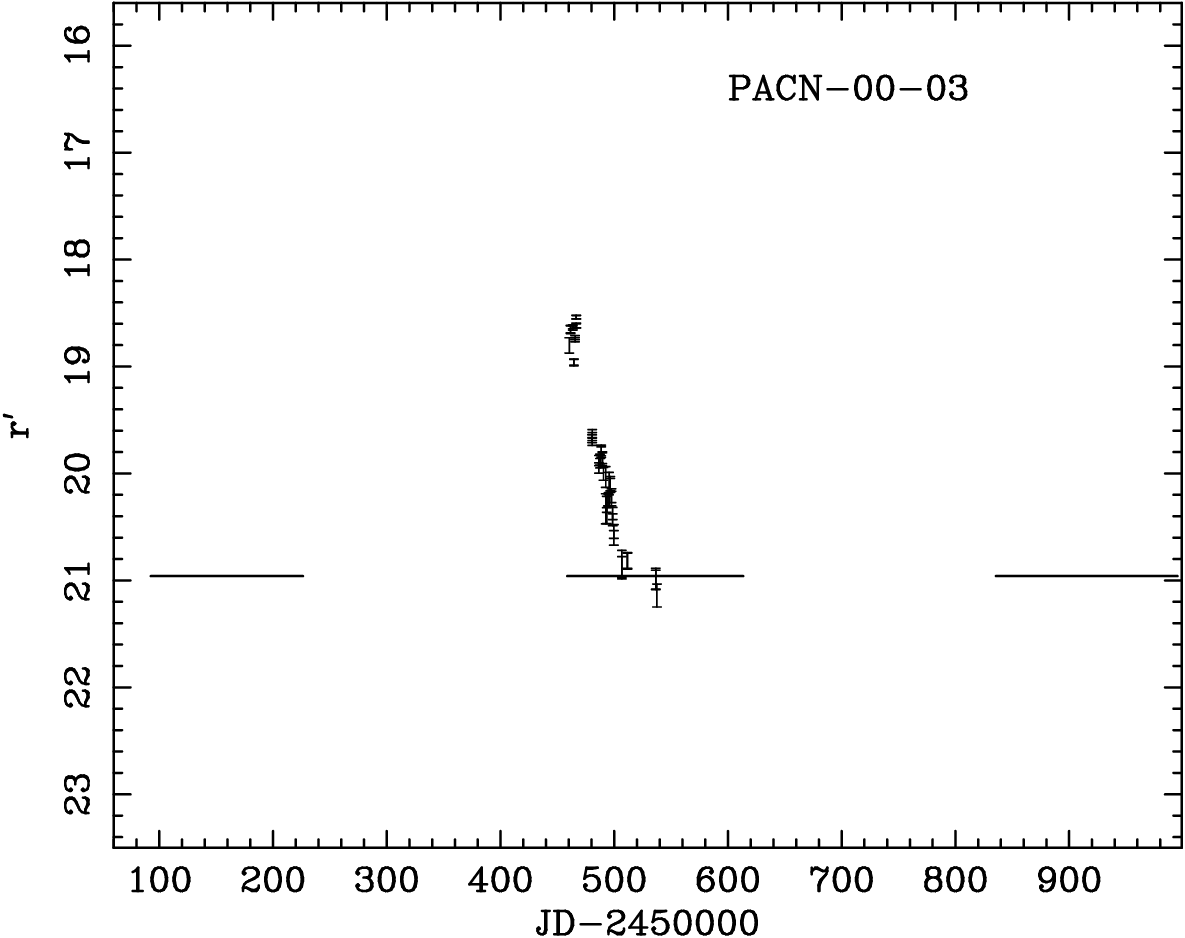}}
\rotatebox{270}{\includegraphics[width=198pt]{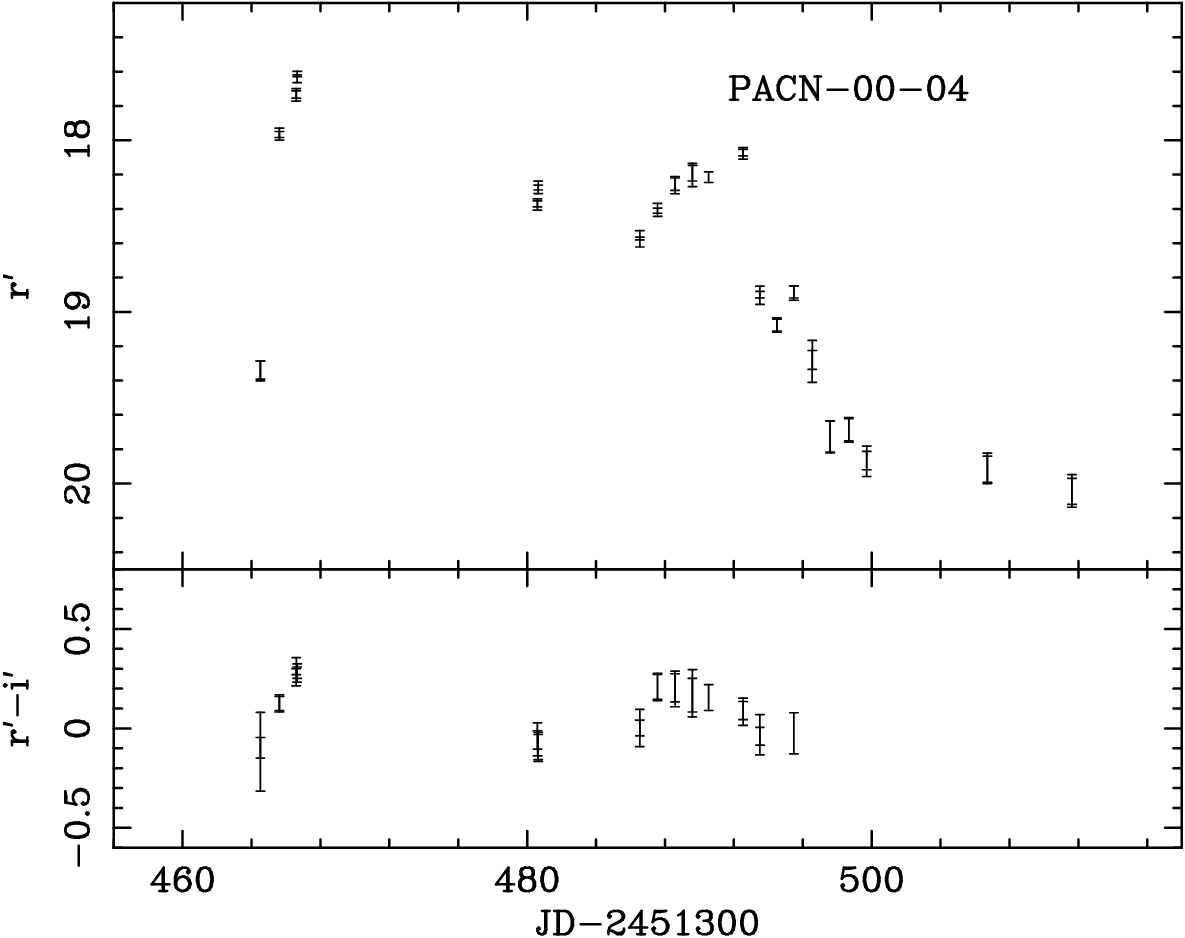}}
\rotatebox{270}{\includegraphics[width=198pt]{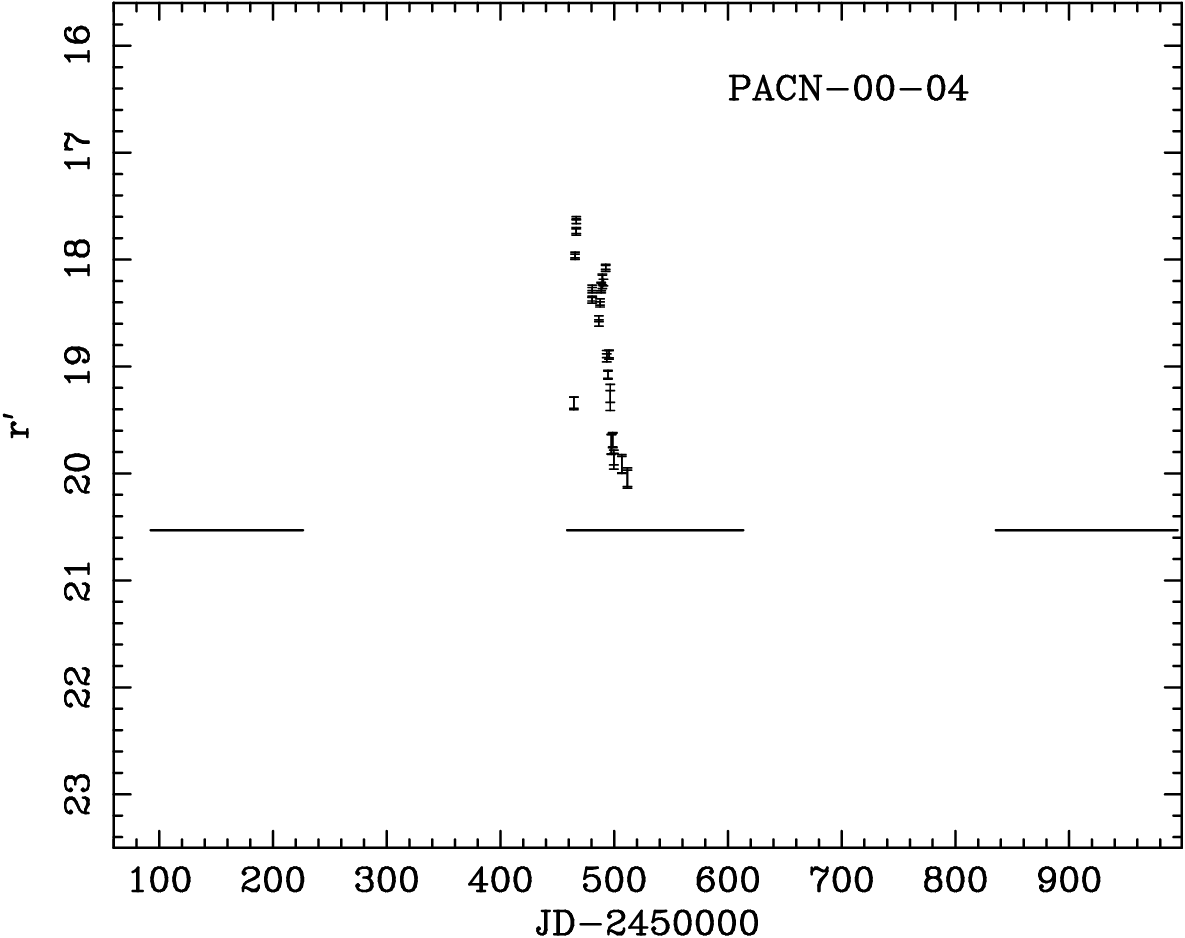}}
\caption{continued.}
\label{mfast-b}
\end{figure*}
\addtocounter{figure}{-1}
\begin{figure*}
\rotatebox{270}{\includegraphics[width=198pt]{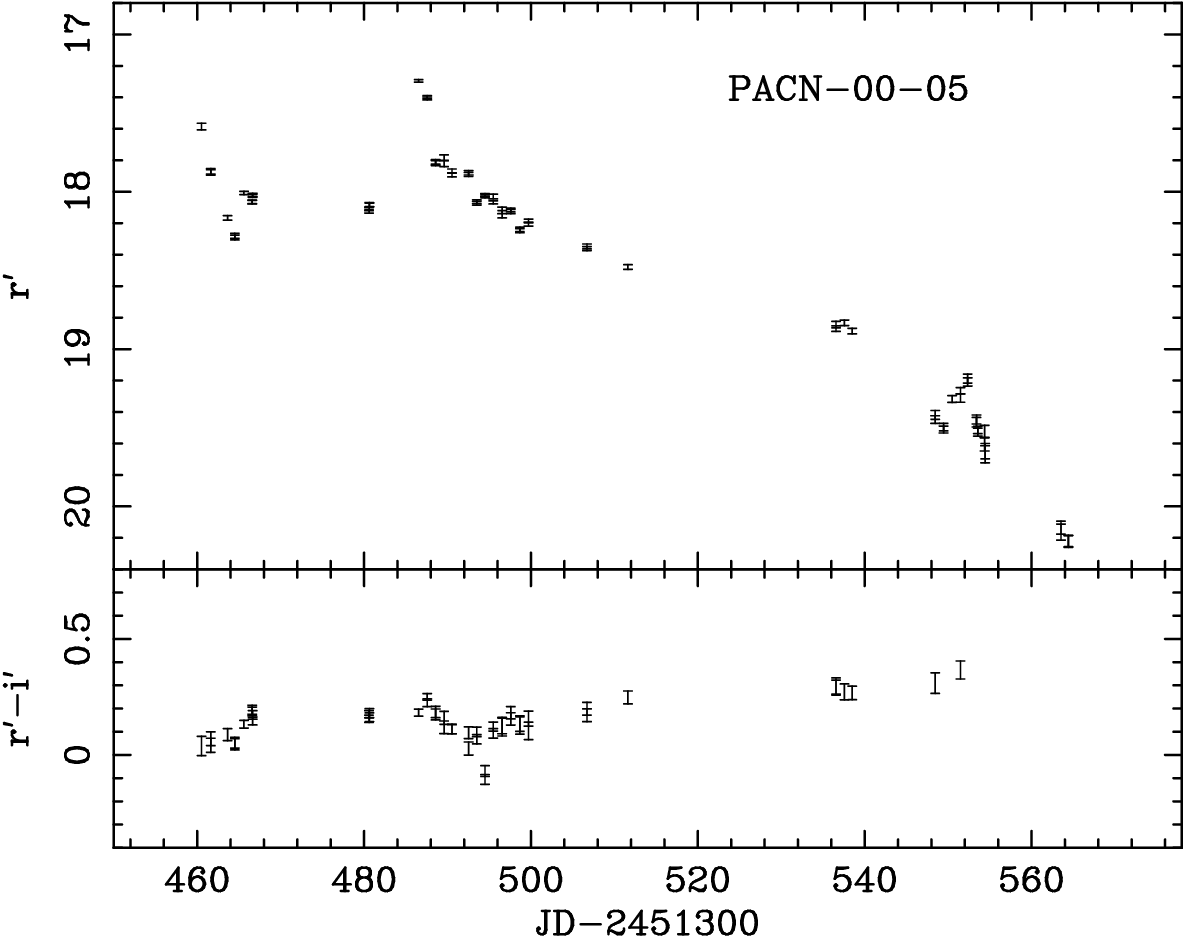}}
\rotatebox{270}{\includegraphics[width=198pt]{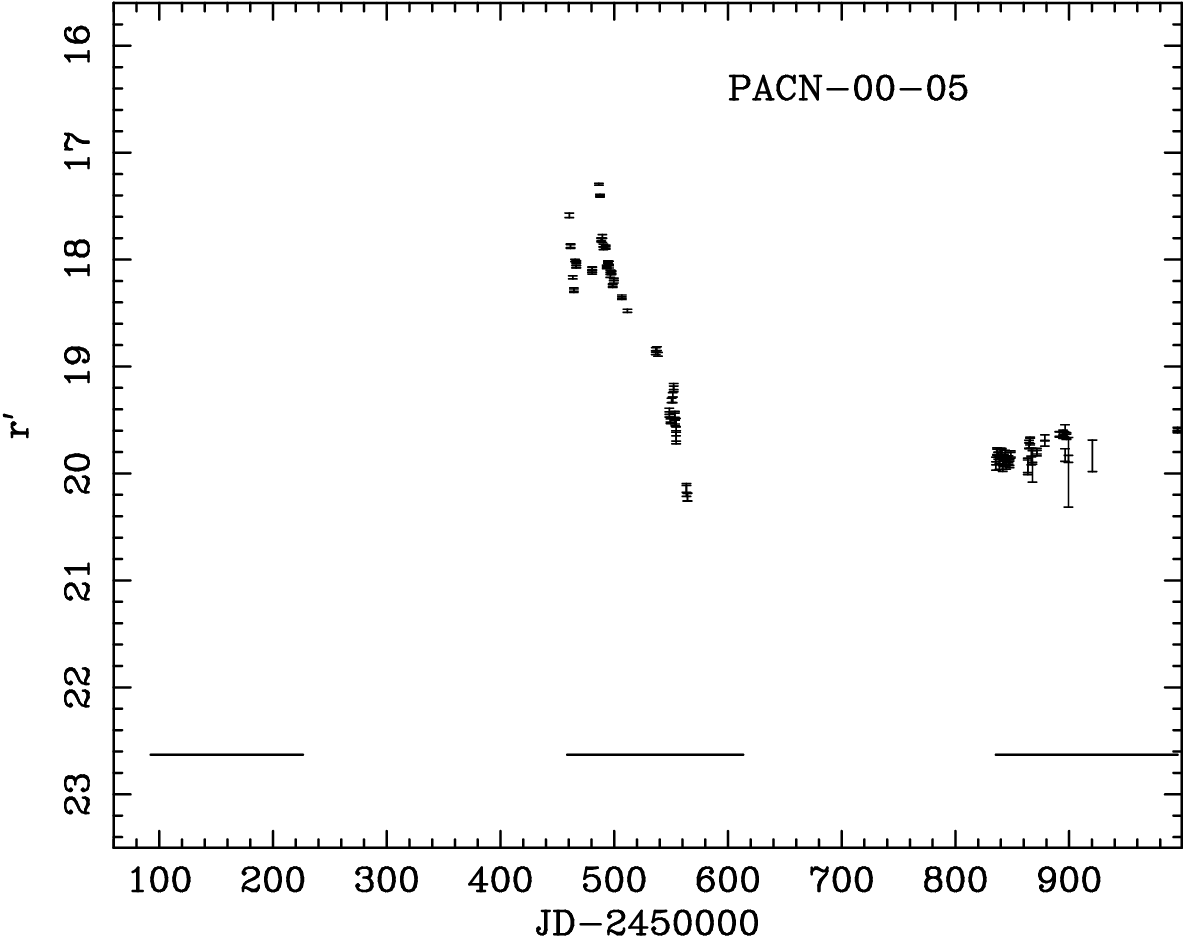}}
\rotatebox{270}{\includegraphics[width=198pt]{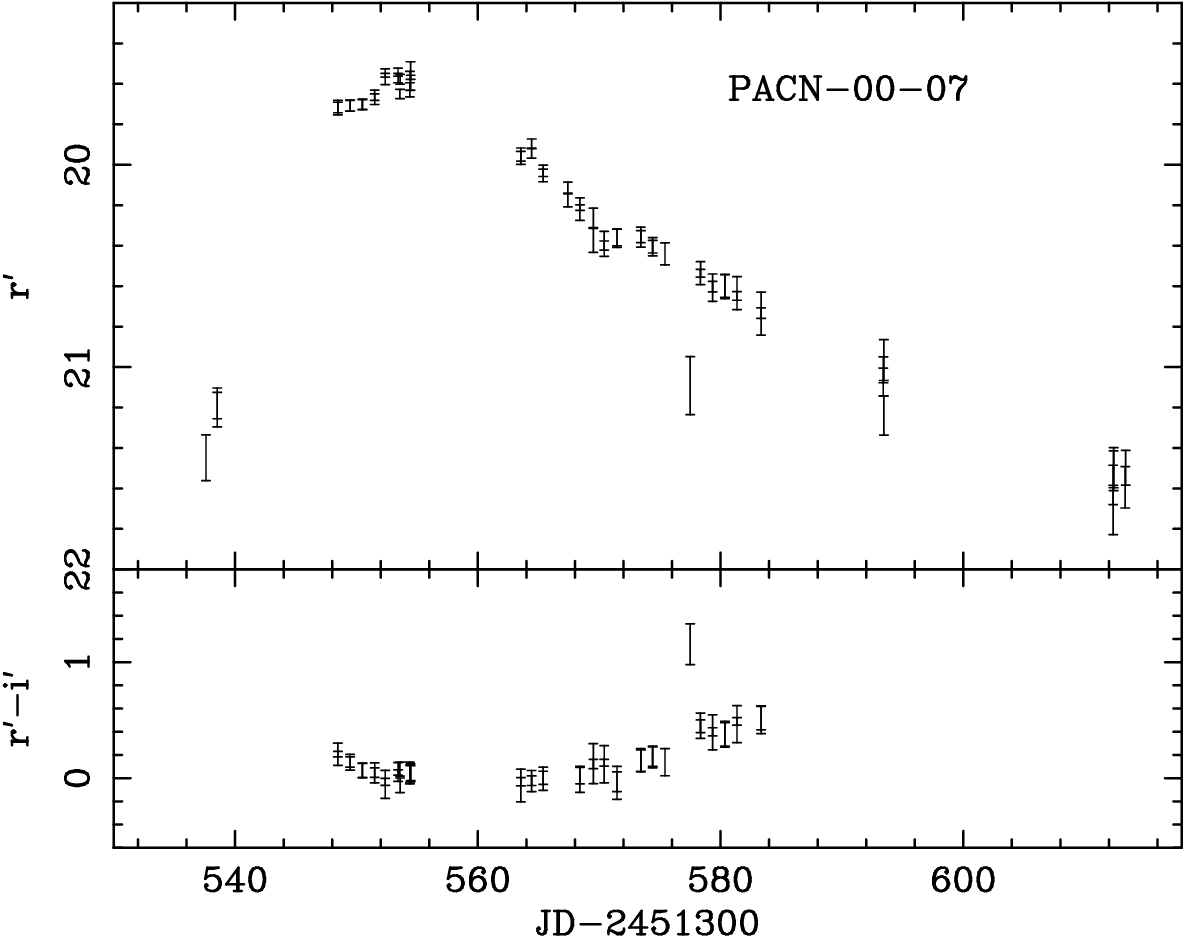}}
\rotatebox{270}{\includegraphics[width=198pt]{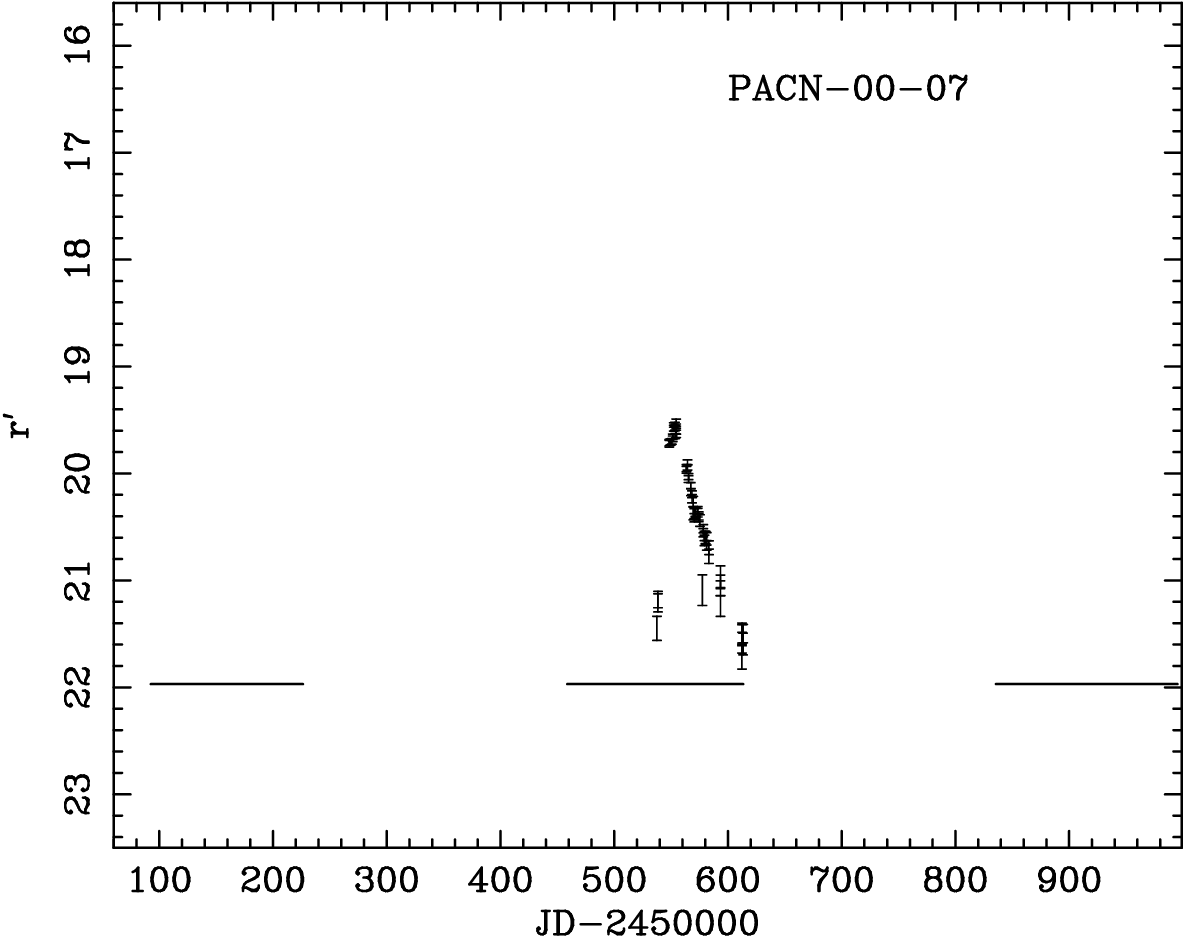}}
\rotatebox{270}{\includegraphics[width=198pt]{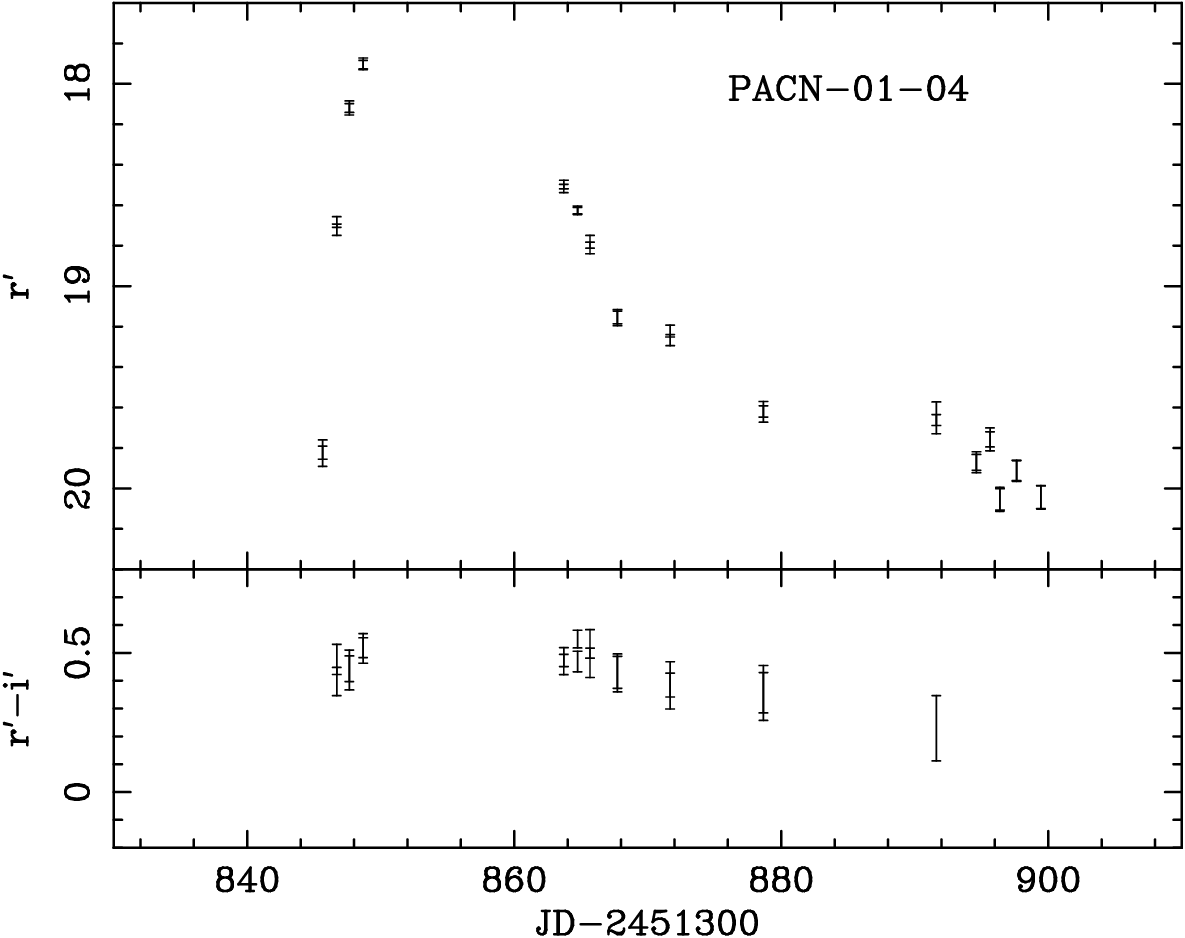}}
\rotatebox{270}{\includegraphics[width=198pt]{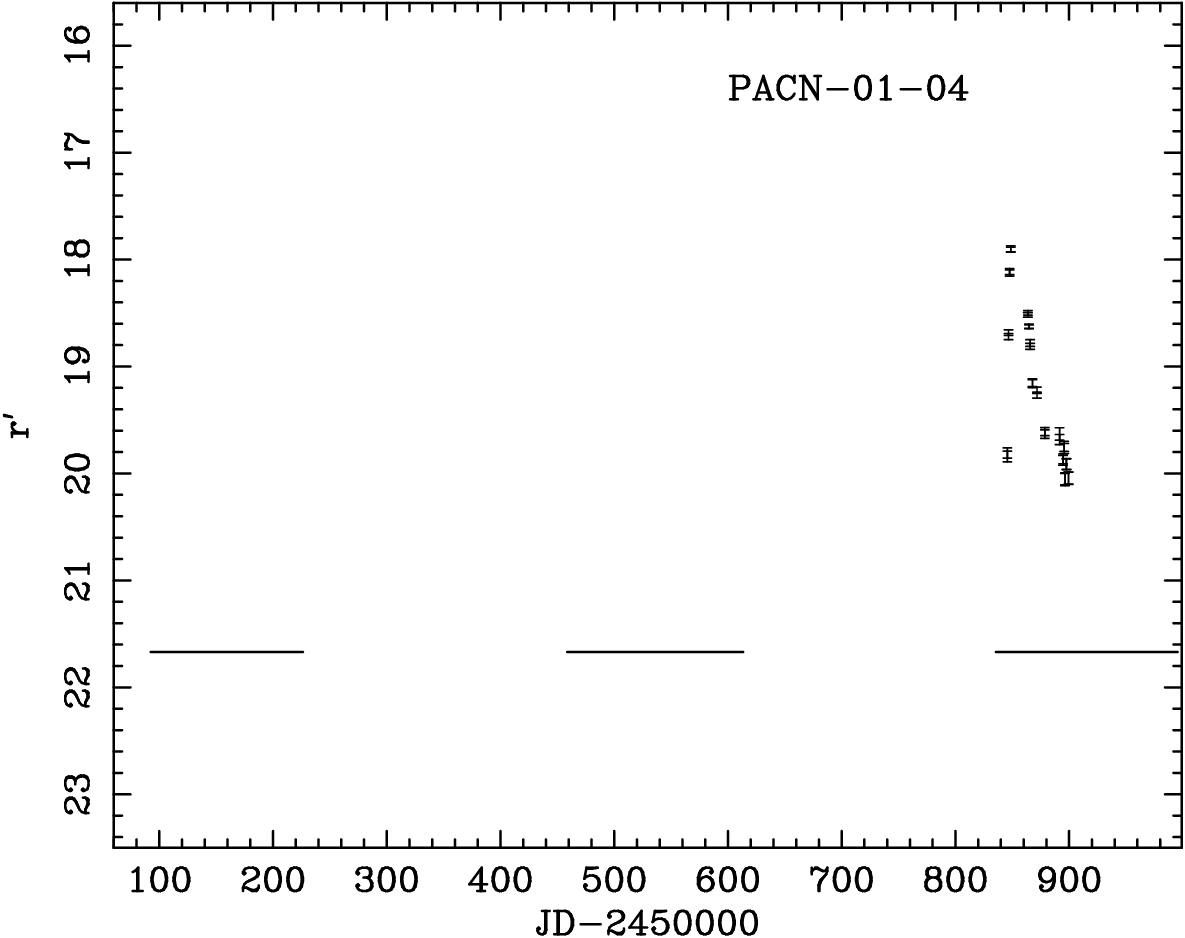}}
\caption{continued.}
\label{mfast-c}
\end{figure*}
\addtocounter{figure}{-1}
\begin{figure*}
\rotatebox{270}{\includegraphics[width=198pt]{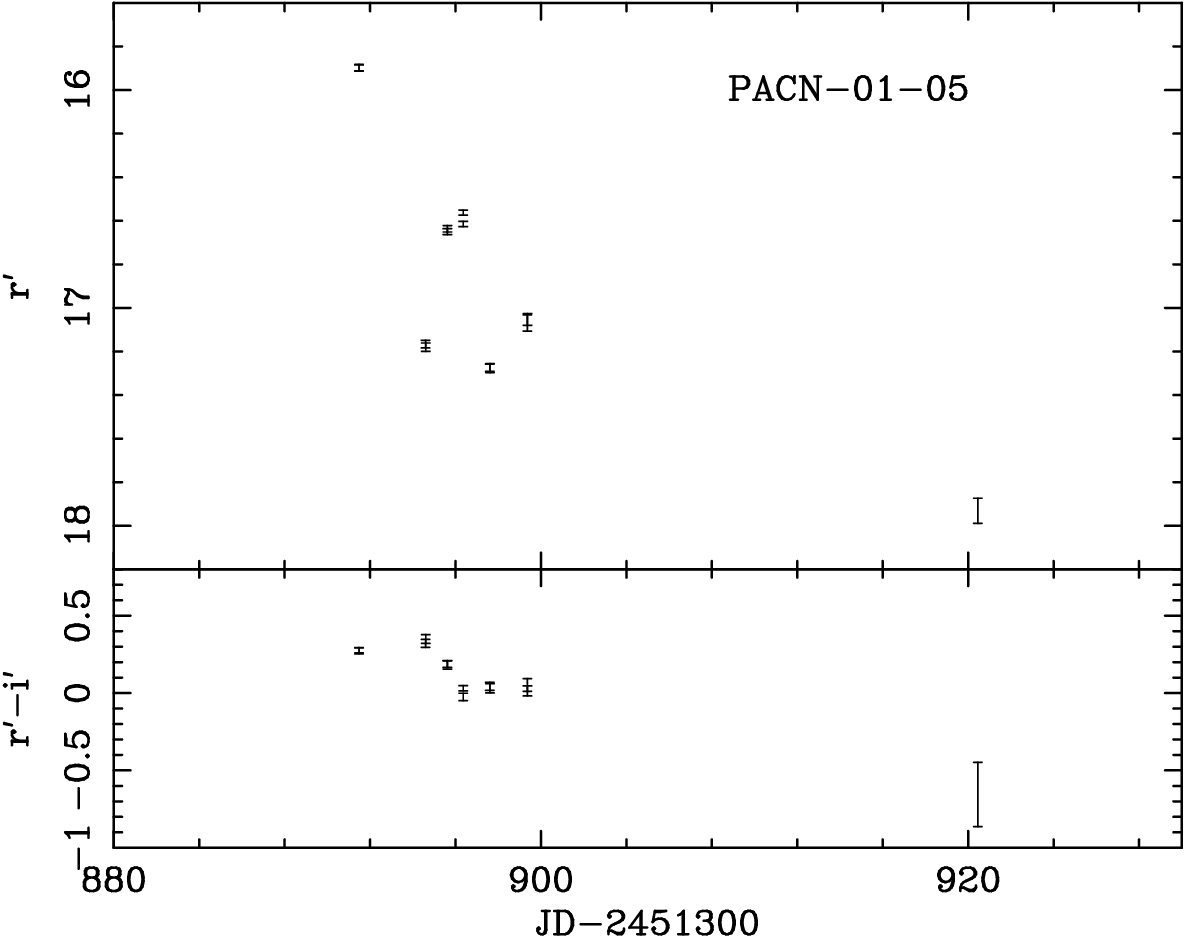}}
\rotatebox{270}{\includegraphics[width=198pt]{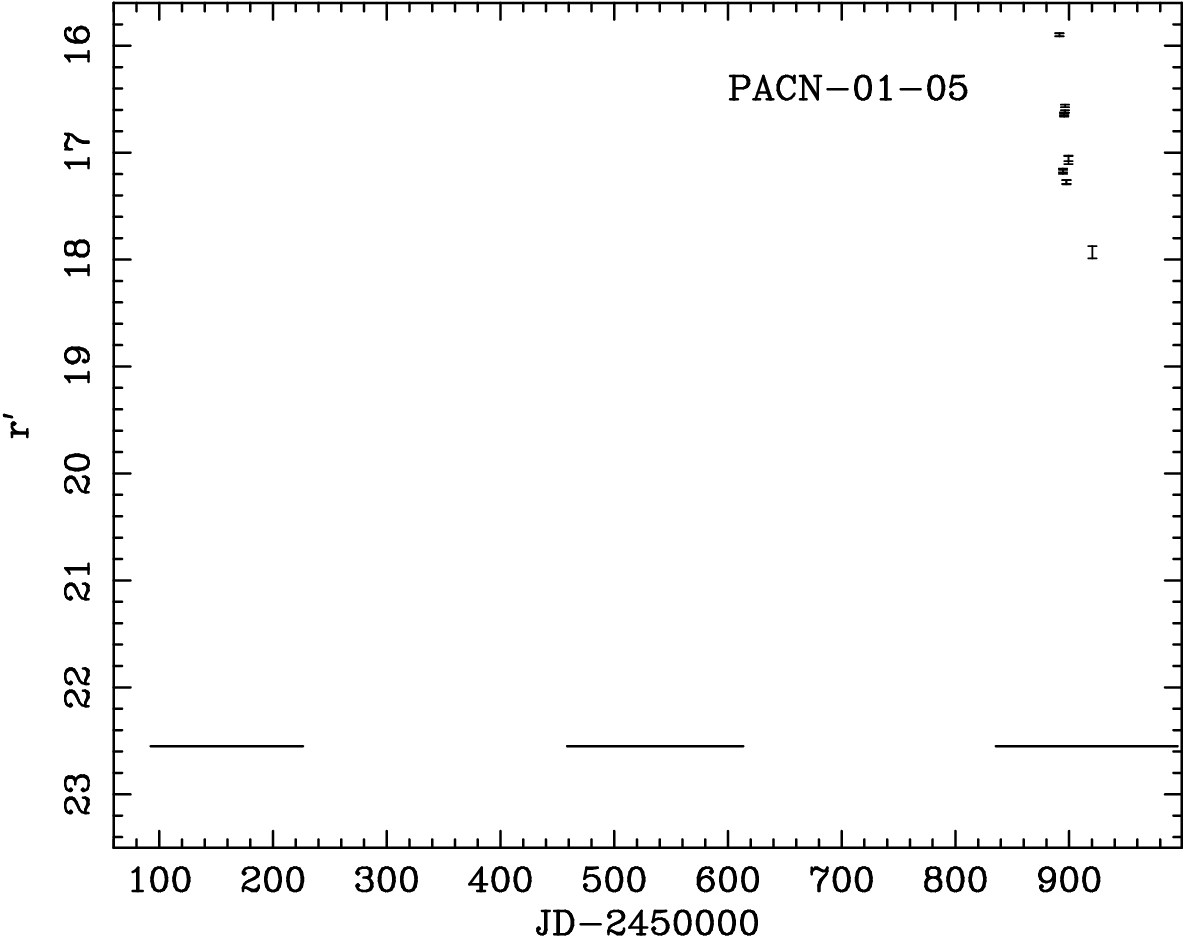}}
\rotatebox{270}{\includegraphics[width=198pt]{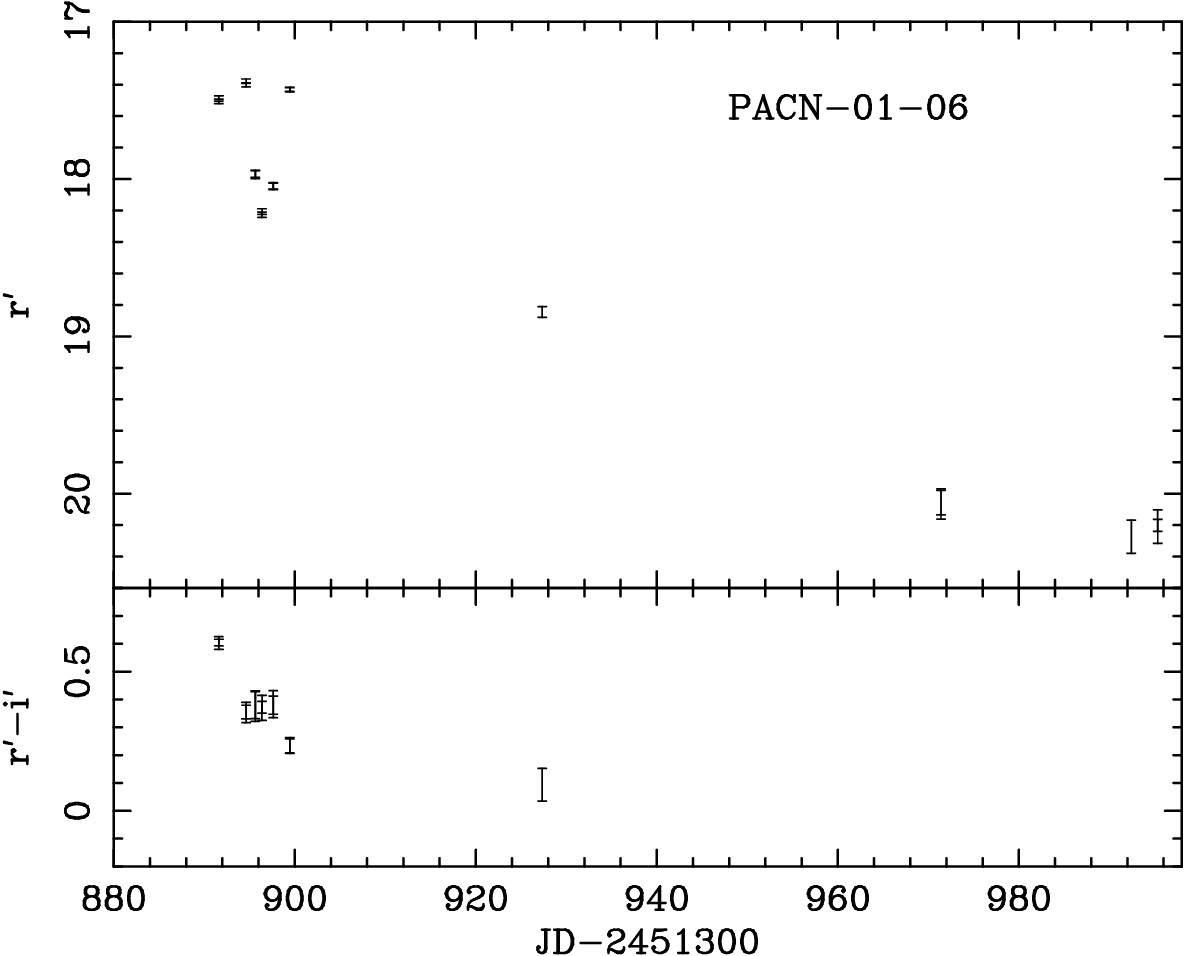}}
\rotatebox{270}{\includegraphics[width=198pt]{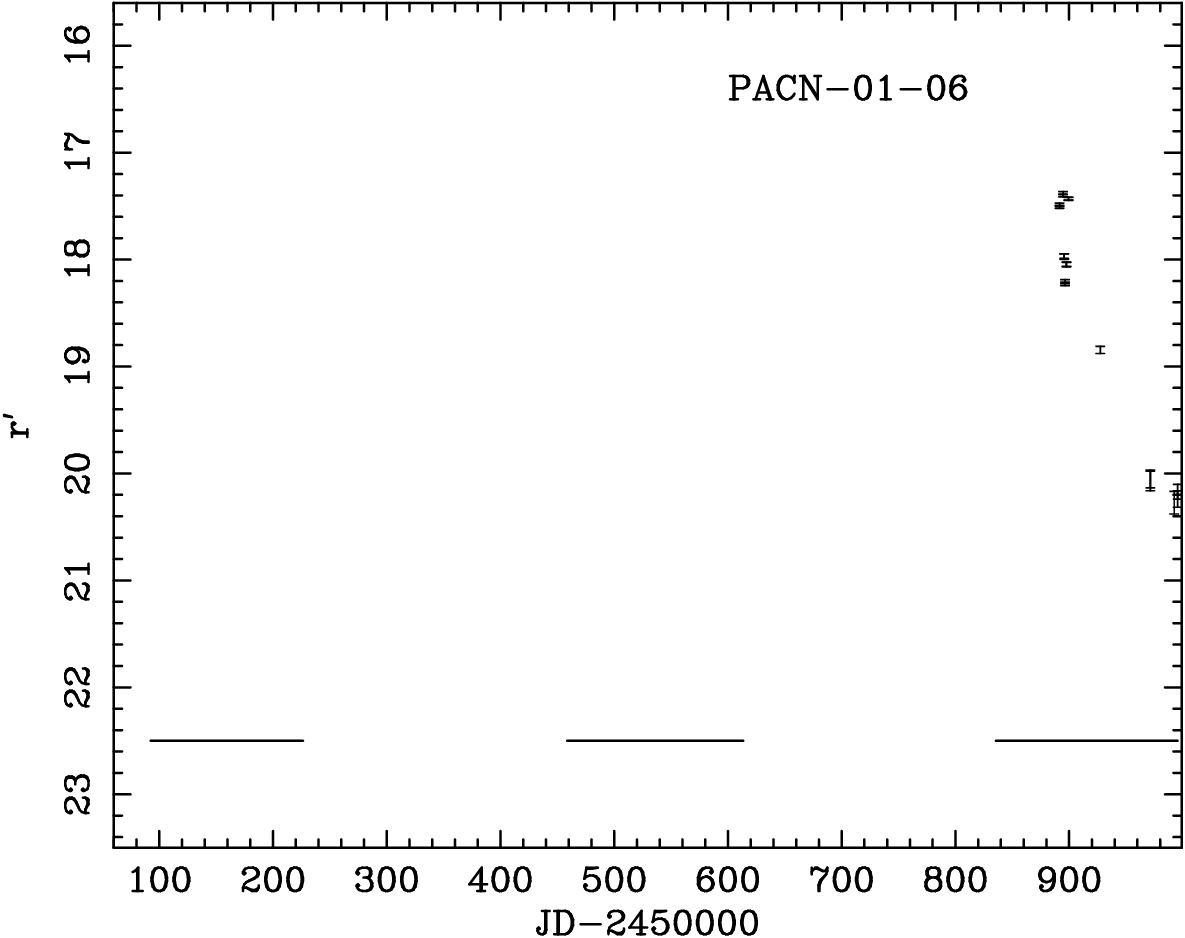}}
\caption{continued.}
\label{mfast-d}
\end{figure*}
We have discovered eleven moderately-fast novae, those with a $dr'/dt$
in the range $0.025-0.07$~mag~day$^{-1}$. Their light-curves are shown in 
Figure~\ref{mfast}.

Five of these novae (PACN-99-06, PACN-00-04, PACN-00-07, PACN-01-04 and
PACN-01-06) were first seen during their final rise phase, with the
remaining novae all appearing to be first observed around or just
after maximum-light.  Two of the moderately-fast novae (PACN-00-04 and
PACN-00-05) exhibit strong oscillations in their light-curves around
maximum light, as is expected for some moderately fast
novae. PACN-00-05 also shows evidence of a large transition phase
minimum between its early and late decline stages, typical of that associated 
with the rapid formation of an optically thick dust shell in the ejecta 
\citep{1989clno.book.....B}. The light-curve of PACN-99-06, contains four points 
in both $r'$ and$g'$ during the final rise phase before reaching a maximum of
$r'=16.17\pm0.01$ and $g'=16.91\pm0.01$.  This CN was first observed
on 7th September 1999 with $r'=17.36\pm0.03$.  When observed again,
just over 24 hours later, it had increased in brightness by 0.9
magnitudes.  This CN was observed two days later at its maximum-light
(11th September), so in the 78 hours prior to maximum-light this CN
had increased in brightness by 1.2 magnitudes.  We were able to follow
PACN-99-06 for about 5 magnitudes below peak.  The light-curve of
PACN-99-06 appears to show that we have observed the pre-nova and
post-nova light for this nova, however this would give this nova a
range of only $\sim5.5$ magnitudes, highly unlikely for any CN.  In
fact the points in the light-curve, all at $r'\sim21$ are contaminants
from a nearby object, whose position is within the errors allowed for
each object (see Section~\ref{detectpipe}).  When PACN-99-06 is
unresolved, at the beginning of the 1st season and for the entirety of
the 2nd and last, the PSF-fitting photometry procedures have
re-centred upon this nearby object -- a relatively faint resolved star
in the M31 field.

The nova PACN-00-04 was observed six times before maximum light, first
on 8th August 2000 with $r'=19.34\pm0.06$, before rising to
$r'=17.61\pm0.03$ just over two days later.  A secondary maximum was
observed at $r'=18.07\pm0.03$ on 1st September.  This nova was only
followed for about 2.5 magnitudes below peak.

The light-curve of PACN-01-04 contains six points before the observed maximum of 
$r'=17.90\pm0.03$ at 4:25 on $27^{th}$ August 2001.  It
was first observed at on $24^{th}$ August at 1.9 magnitudes below peak and was 
followed for about two magnitudes after its maximum light.
PACN-00-05 was first observed on 4th August 2000, during the second
season Its maximum observed brightness was $r'=17.58\pm0.03$. A
secondary maximum of $r'=19.59\pm0.03$ occurred about 500 days after
maximum light following a transition phase.  Unfortunately the
transition phase occurred between the end of the second season and the
start of the third, so no information is available for this nova
during this phase. The nova was followed through two magnitudes before
the end of the second season.

The CN PACN-01-06 was also discovered by the Naini~Tal microlensing
group \citep{2004A&A...415..471J}, and was designated CN~NMS-2 by them.

\begin{figure*}
\rotatebox{270}{\includegraphics[width=190pt]{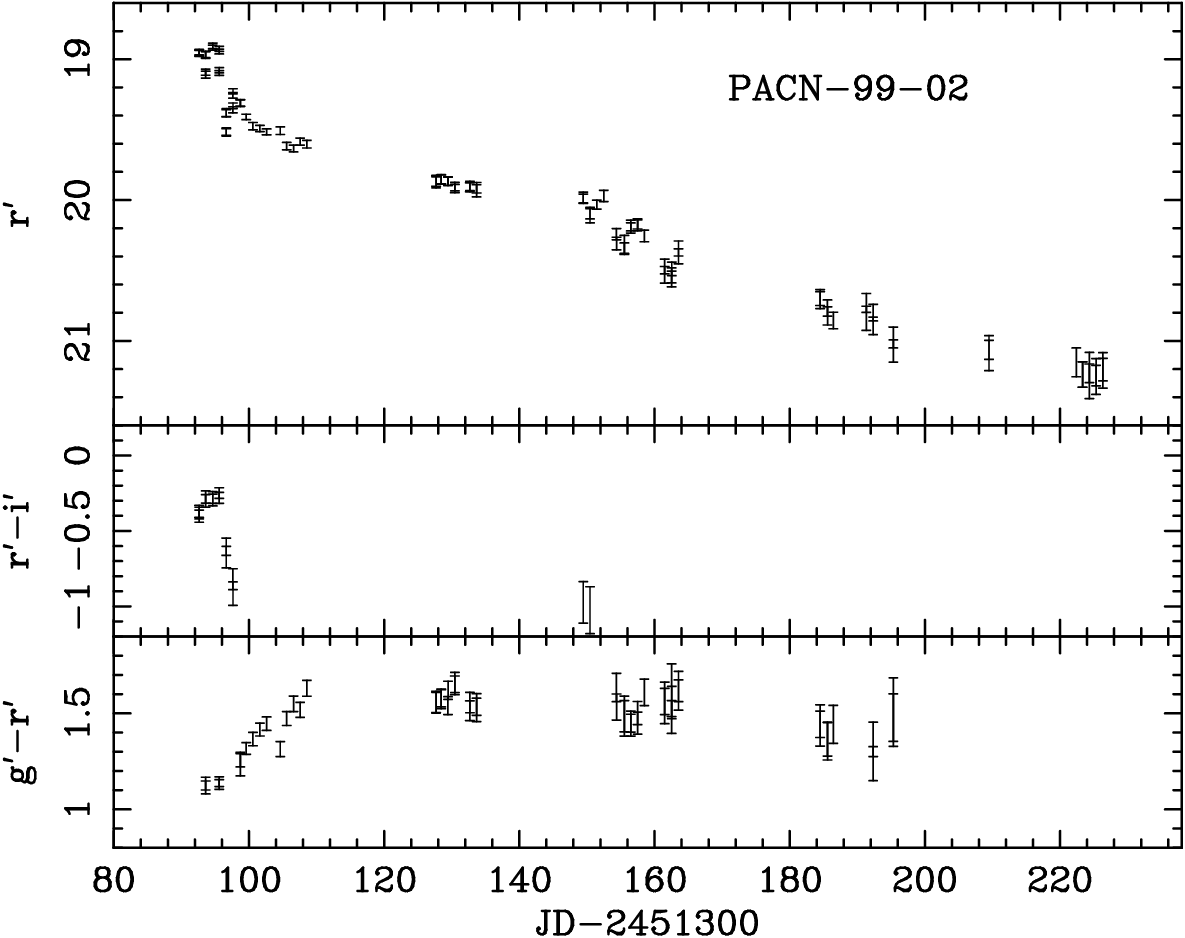}}
\rotatebox{270}{\includegraphics[width=190pt]{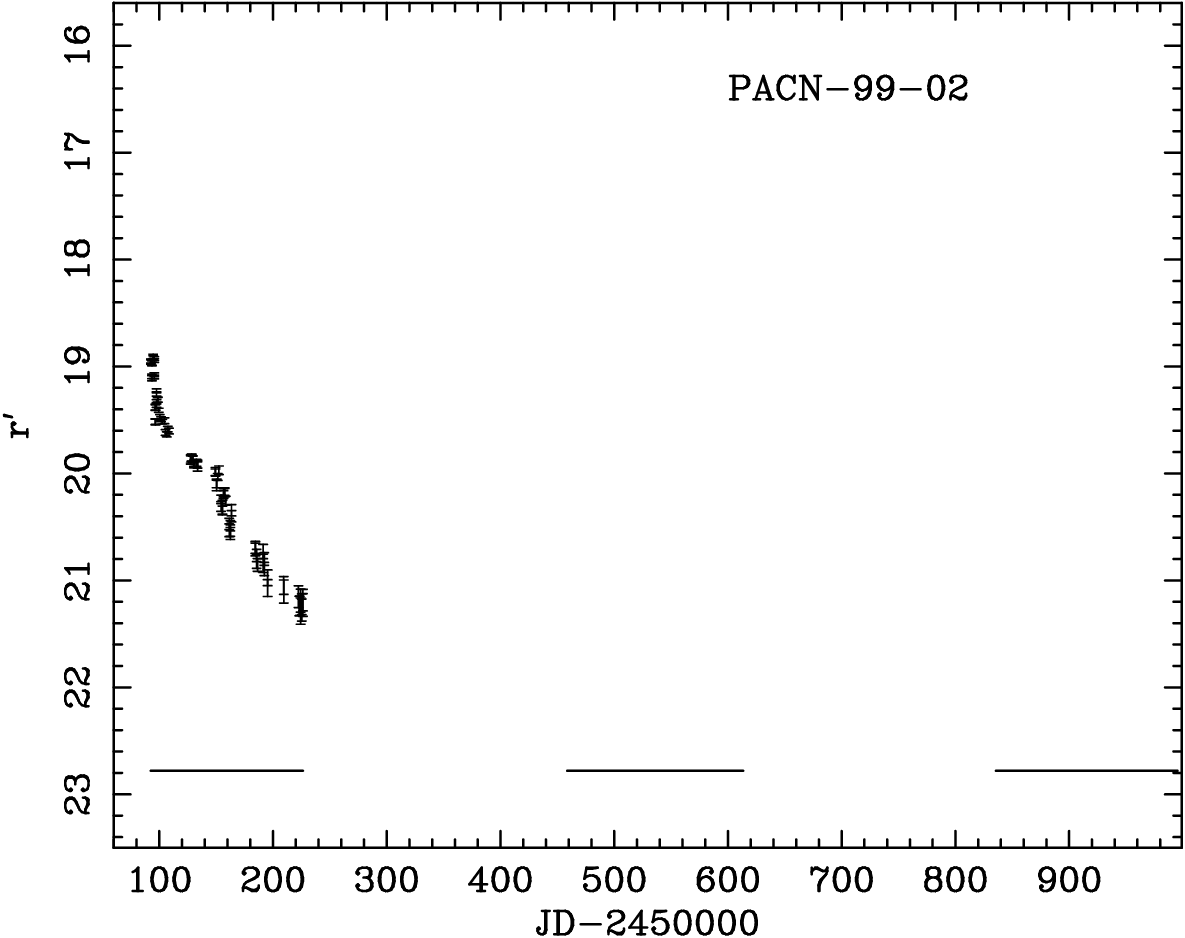}}
\rotatebox{270}{\includegraphics[width=190pt]{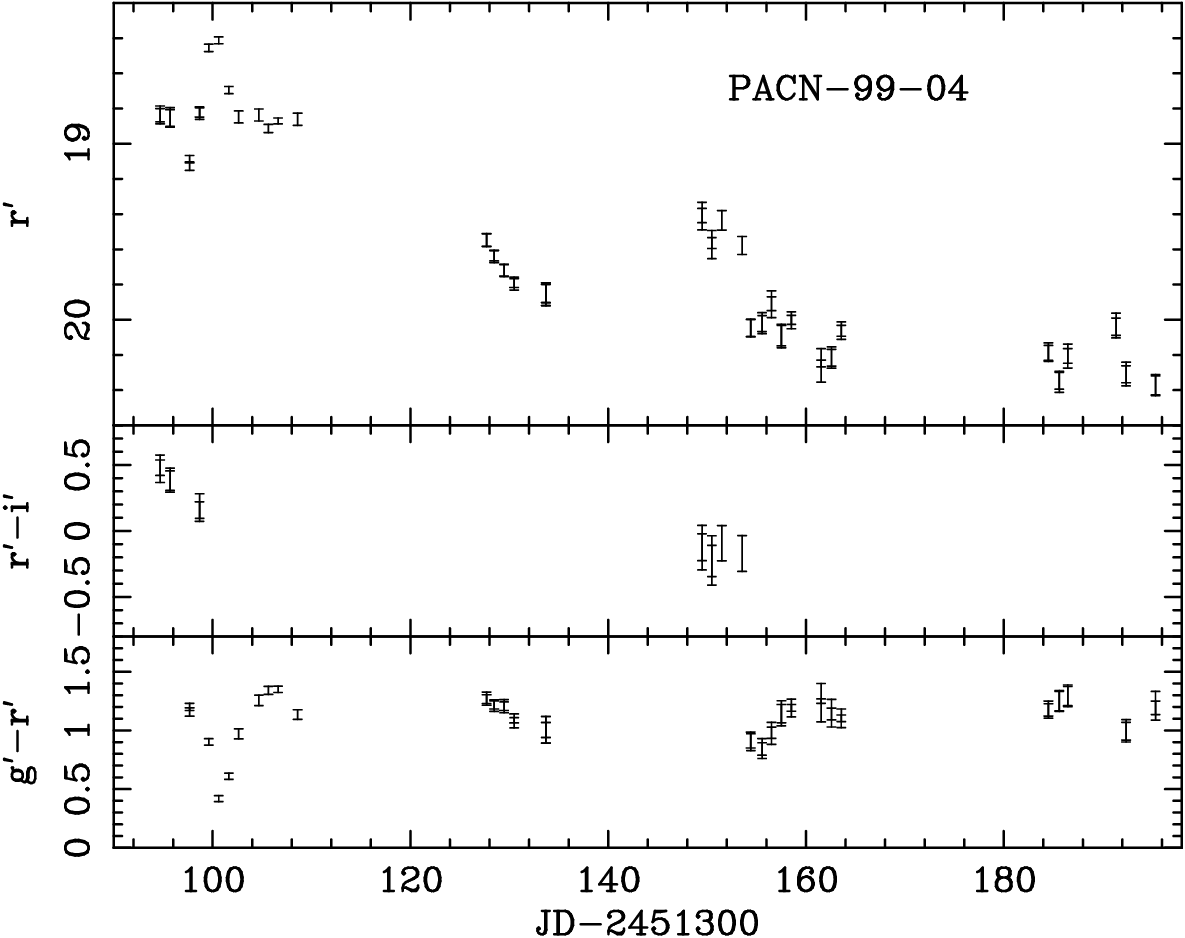}}
\rotatebox{270}{\includegraphics[width=190pt]{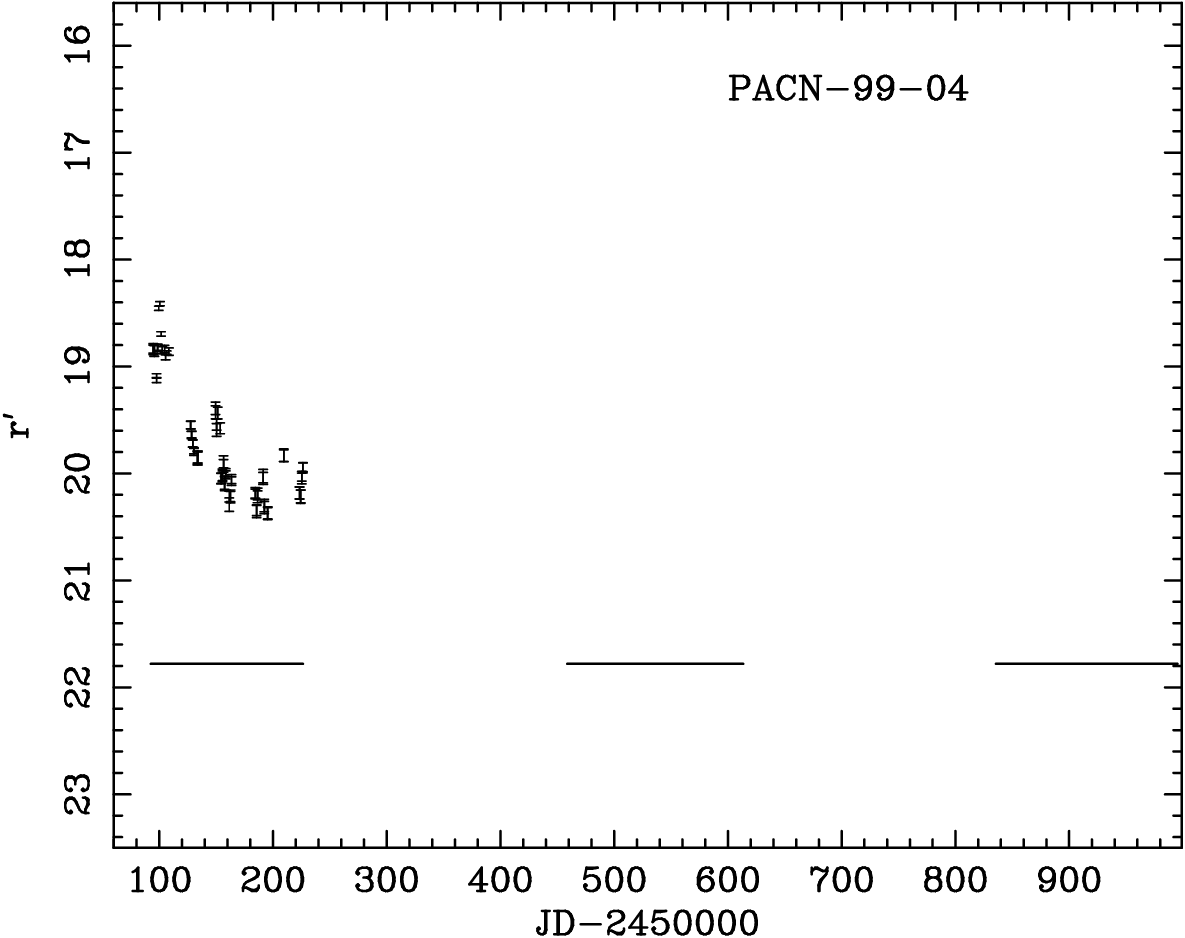}}
\rotatebox{270}{\includegraphics[width=190pt]{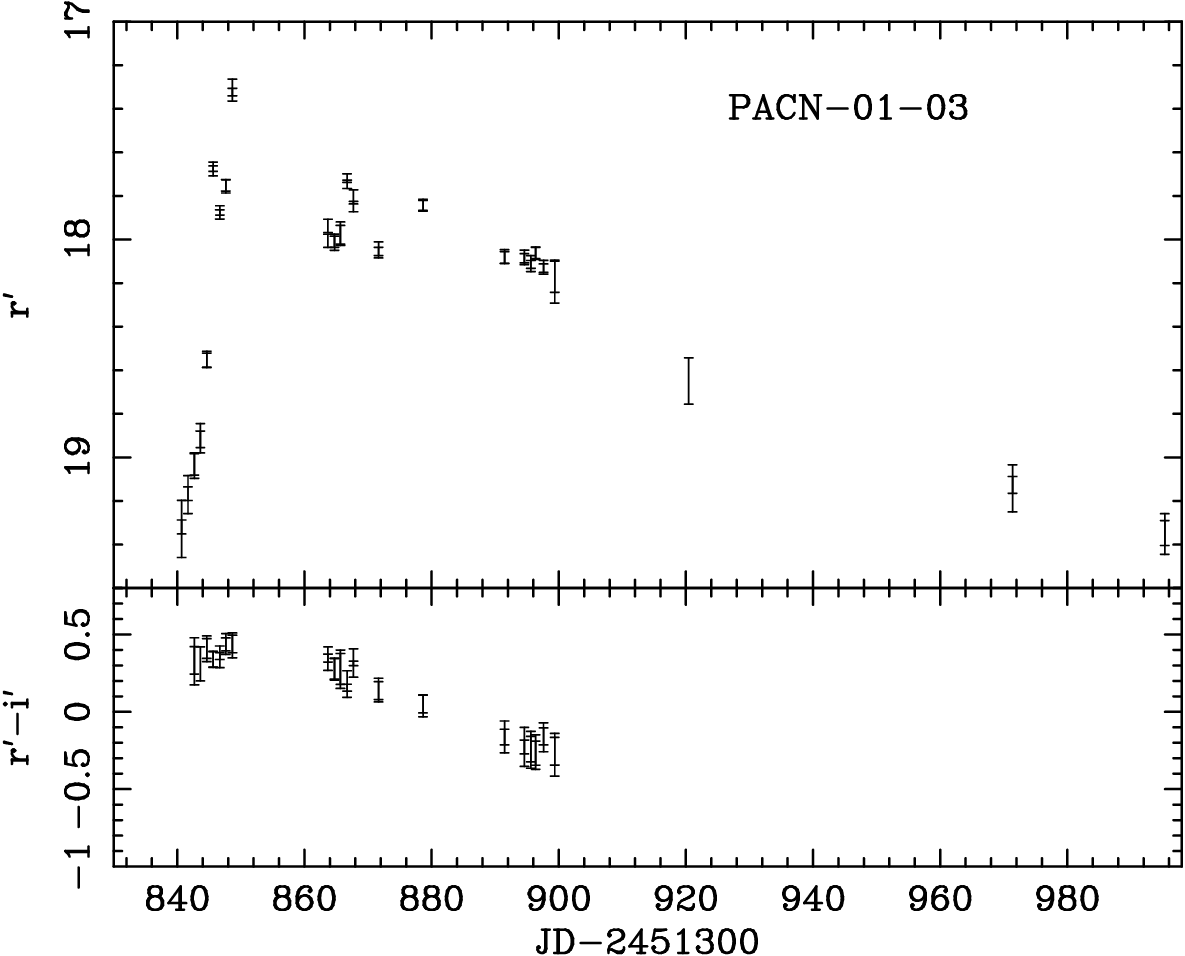}}
\rotatebox{270}{\includegraphics[width=190pt]{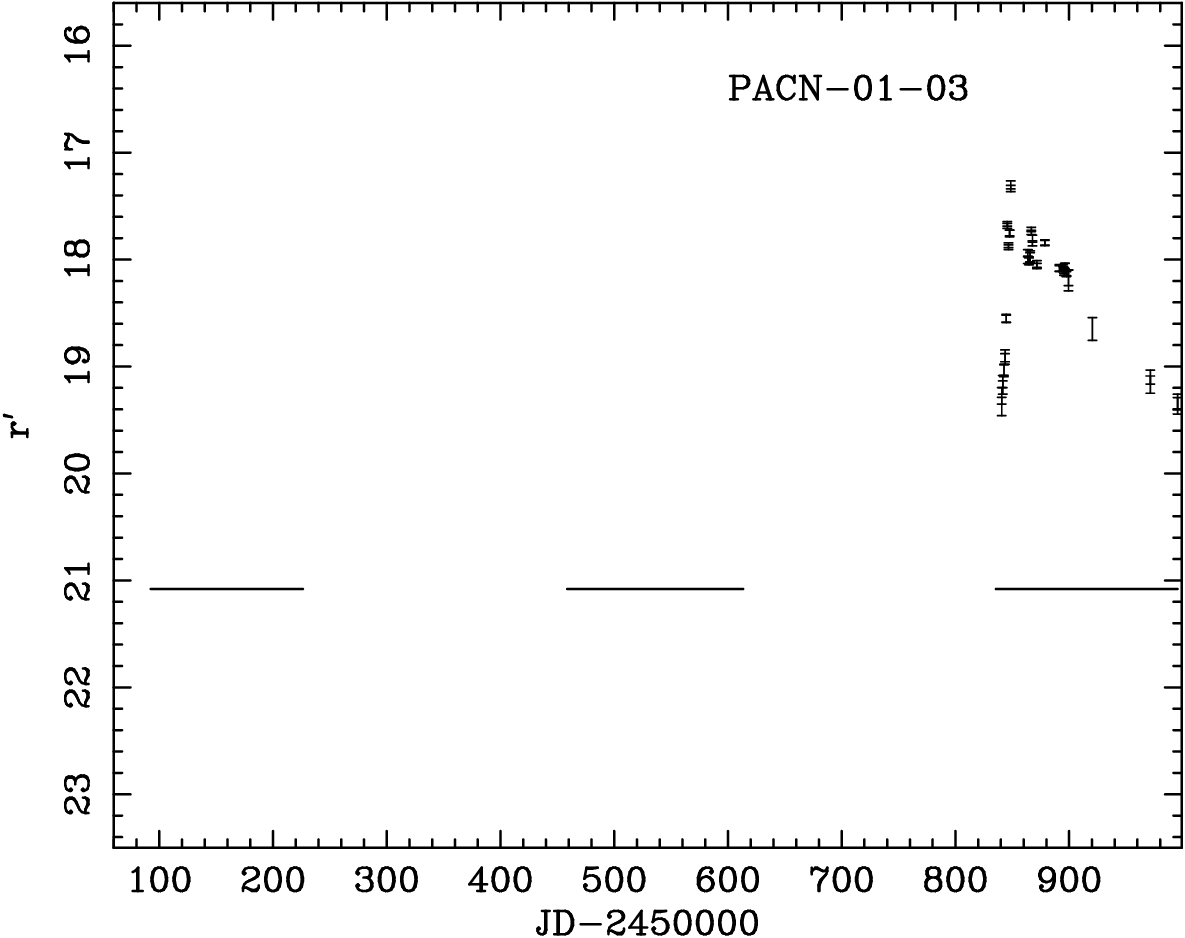}}
\caption{Slow novae in the POINT--AGAPE dataset. The panels are as for Figure~\ref{vfast}.}
\label{slow}
\end{figure*}

\begin{figure*}
\rotatebox{270}{\includegraphics[width=190pt]{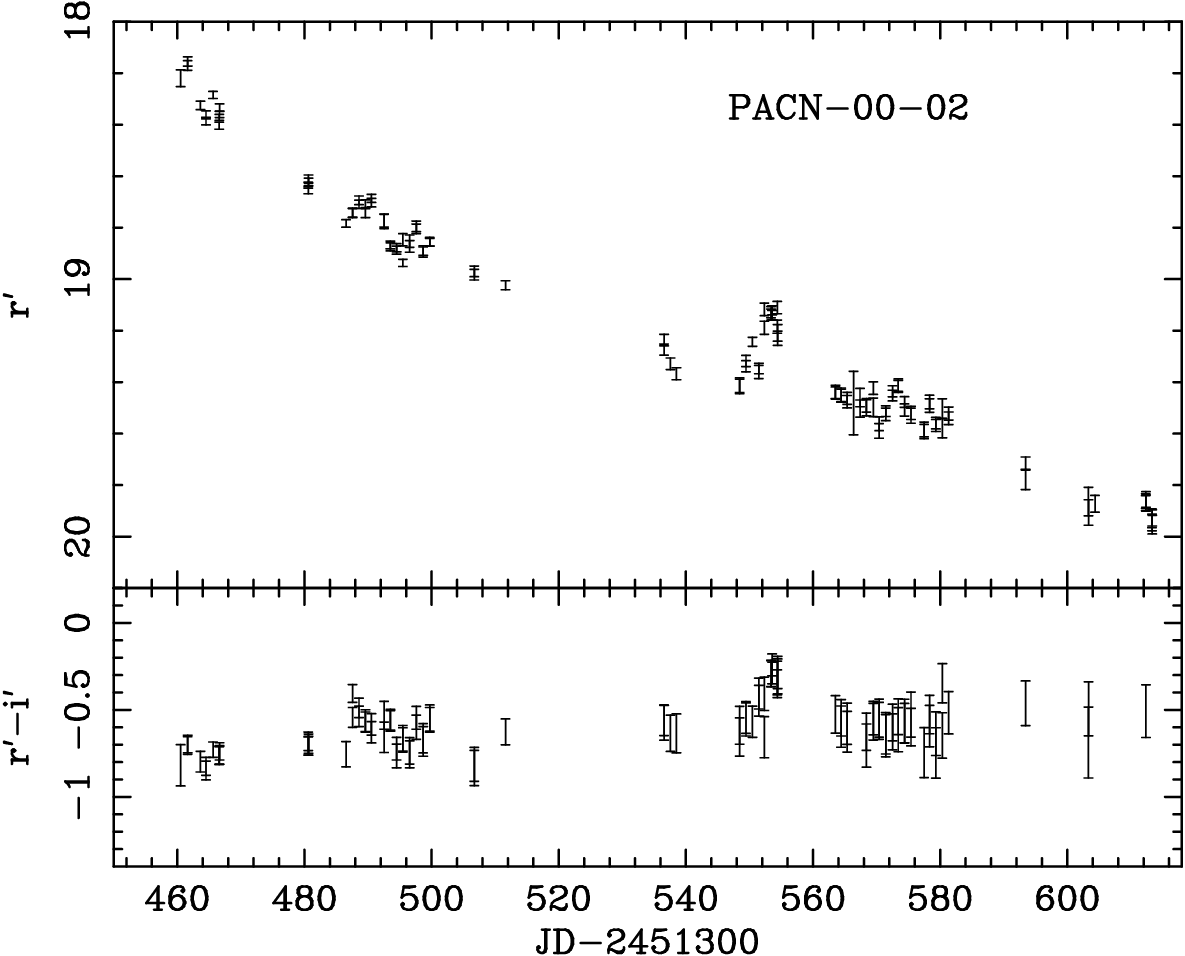}}
\rotatebox{270}{\includegraphics[width=190pt]{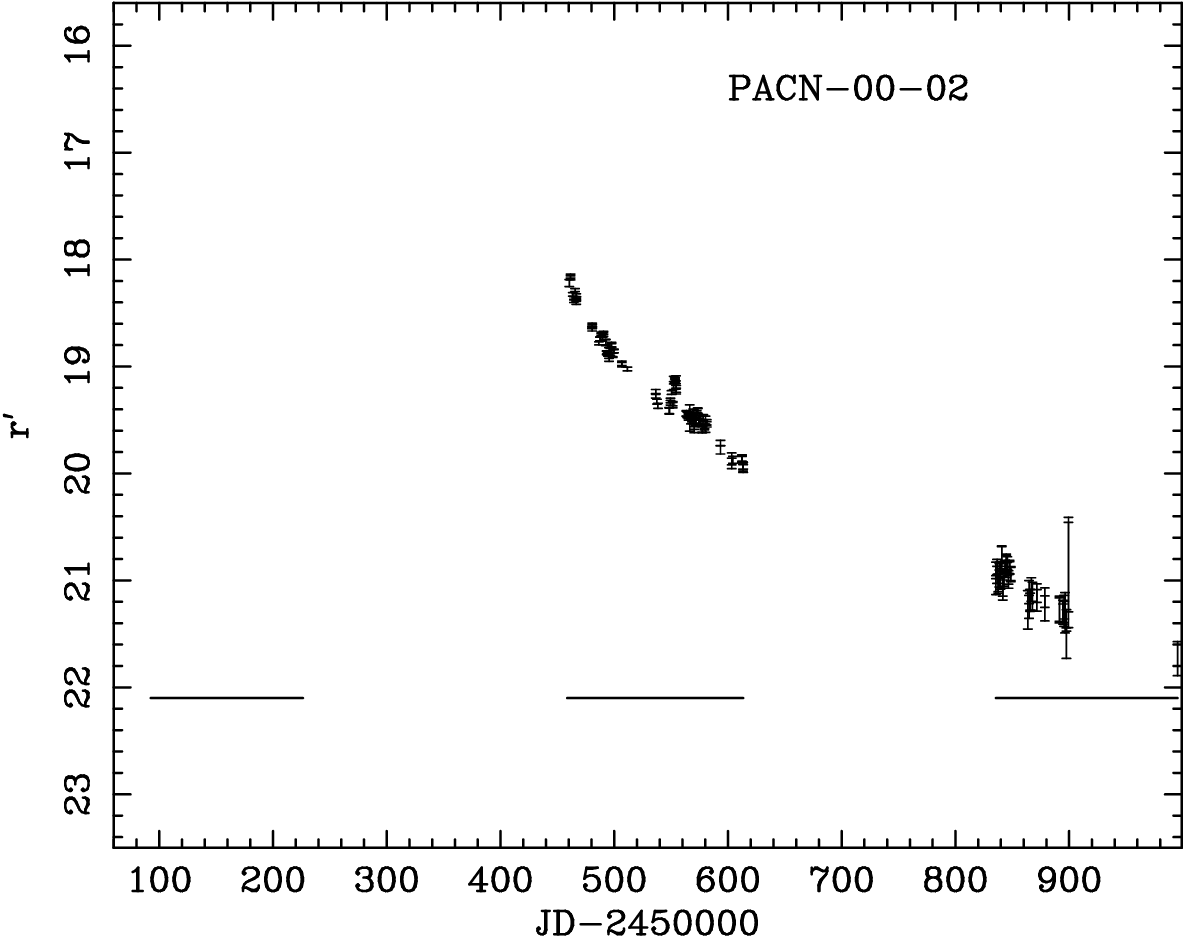}}
\rotatebox{270}{\includegraphics[width=190pt]{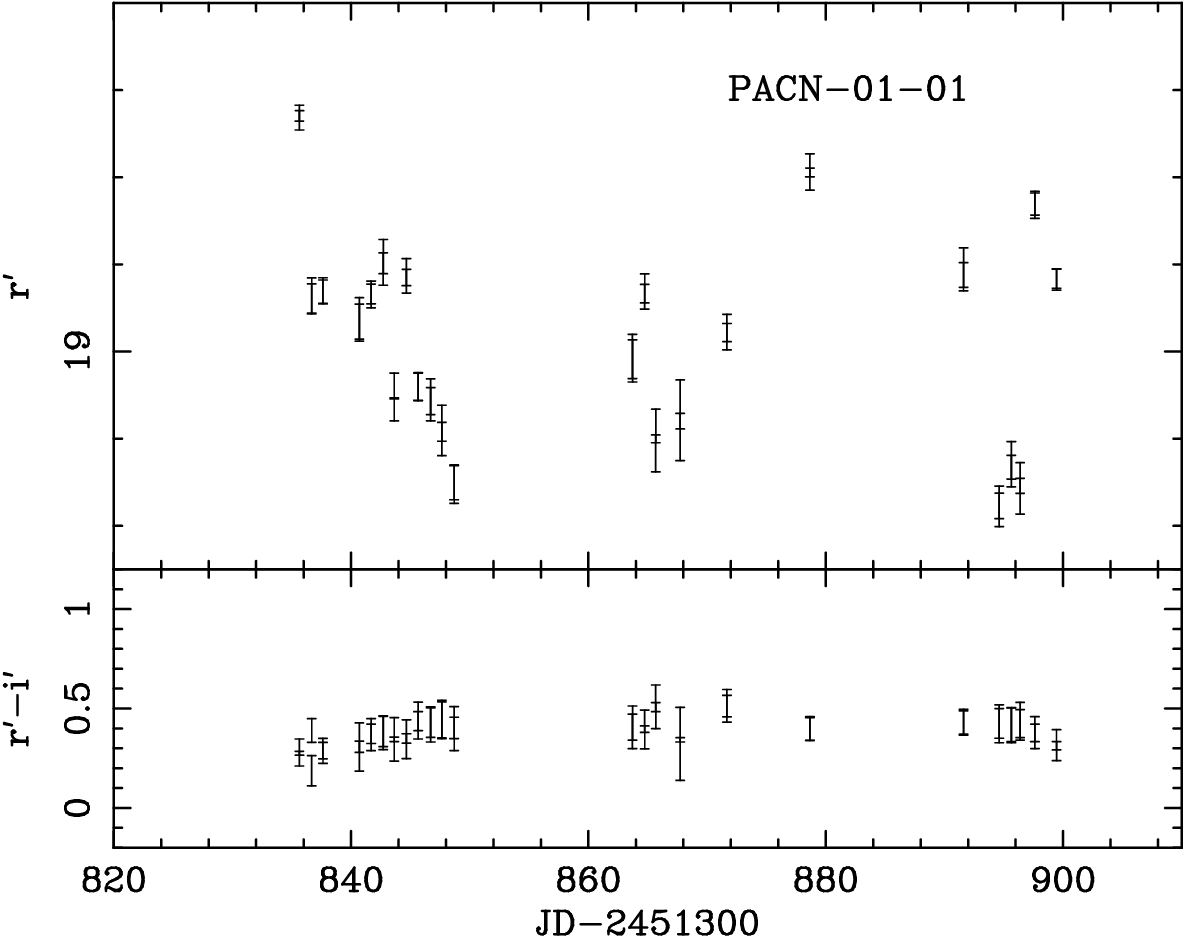}}
\rotatebox{270}{\includegraphics[width=190pt]{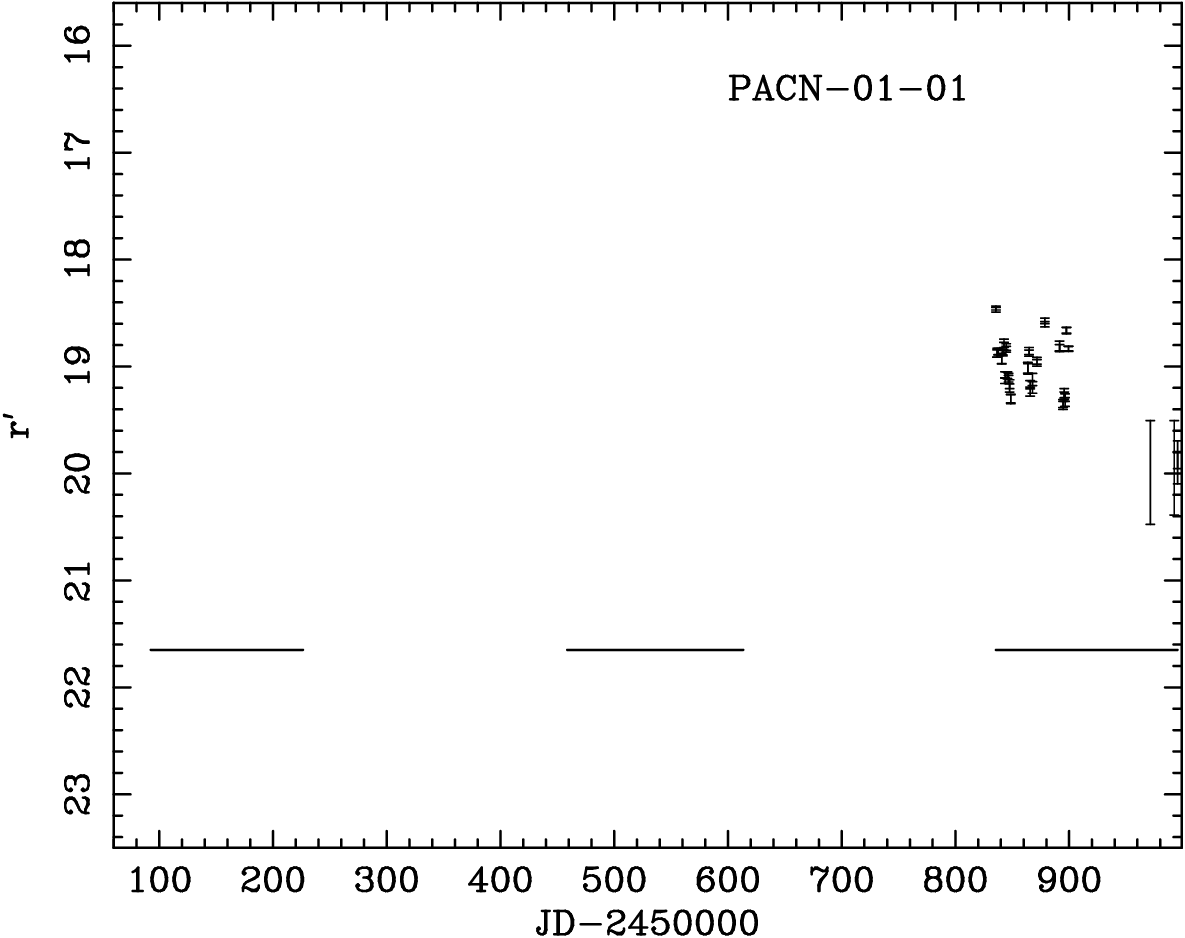}}
\caption{Very slow novae in the POINT--AGAPE dataset. The panels are as for Figure~\ref{vfast}.}
\label{vslow}
\end{figure*}

We are less certain about the classification of PACN-00-07 as a CN. At
first glance it looks very much like that of a CN. The rate of decline
indicates that it is a moderately fast nova. However with a maximum
magnitude of $r'=19.53\pm0.04$ it is much fainter than the other ten
moderately fast CNe which all have maximum light in the range
$r'=15.9$ to $17.9$. Interestingly, PACN-00-07's colour evolution
appears to be more like that of a Mira than of a CN (see
Figure~\ref{mira} for comparison), but its position on the
colour-magnitude diagram (Figure~\ref{colmag}) is much nearer to the
CNe group than that of the other objects (see
Section~\ref{borderl}). Perhaps tellingly, it is the object which lies
closest to the dividing line between the groups in a position which is
suggestive of a significant extinction effect ($\sim 1-2$ magnitudes
in $r'$). It is possible that PACN-00-07 is a highly extinguished nova
or that it is a CN whose maximum light occurred during the gap between
the 1st and 2nd season, and we are just observing a local maximum in
the light-curve.  Conceivably, this object could be a recurrent CN or
perhaps an X-ray nova, though these properties would not necessarily
result in a fainter maximum than expected for ``regular'' CN, and a
search for possible X-ray counterparts in the Nasa Extragalactic
Database (NED) reveals no matching candidates. In our view the most
probable explanation is that the light from this CN is attenuated by
dust.

\subsection{Slow novae}

Three slow novae were discovered, with a $dr'/dt$ of $0.013-0.024$
mag~day$^{-1}$, and are displayed in Figure~\ref{slow}.  Two of these
slow CNe, PACN-99-04 and PACN-01-03, were first observed during their
final rise phase.  However, as slow novae are generally fainter than the fast
novae (at maximum light), it was not possible to follow either
PACN-99-02 or PACN-99-04 into their transition stage. PACN-01-03,
although somewhat brighter, occurred at the end of the third season.

PACN-99-04 was first observed on 4th August 1999, the first epoch of
the 1st season, with $r'=18.84\pm0.04$.  The maximum light was observed 
on 9th August with $r'=18.41\pm0.03$.  The nova was followed for
about 2 magnitudes below peak.

The CN PACN-01-03 was observed seventeen times throughout its final
rise phase.  This nova was first observed with $r'=19.27\pm0.08$ and
rose steadily for 8 days to its observed maximum of
$r'=17.30\pm0.04$ on 27th August 2001.  We were only able to follow
this nova during its initial decline phase for around two magnitudes
before the end of the third season.

\subsection{Very slow novae}

As shown in Figure~\ref{vslow}, we discovered two very slow CNe. Very slow
CNe take $151-250$ days to decrease by two magnitudes from maximum
light.

The maximum in the $r'$-band light-curve of PACN-00-02 was observed at
the beginning of the second season on 4th August 2000 and the nova was still
clearly visible by the end of the second season on 3rd January 2001.  However, 
it had become unresolved by the beginning of the third.  This CN reached an
observed maximum light of $r'=18.15\pm0.03$ and had diminished by only 1.8
magnitudes by the end of the second season, 150 days later, with an estimated
$dr'/dt\simeq0.01$ mag day$^{-1}$. PACN-00-02 has a relatively smooth
light-curve, except for a feature about 100 days after maximum light in which
the nova brightened by about a third of a magnitude, before continuing to
decline again.

PACN-01-01 is an interesting
object, though we have some reservations over its classification as a nova.  It
was
not possible to sample enough of PACN-01-01's light-curve to make a 
reliable measurement of $dr'/dt$ as this nova remained around maximum-light for
the majority of the time that it was observed. Its speed class, if it is a CN,
is therefore uncertain.
However, it has passed all of our selection criteria.  The 3rd season data are
similar in some respects to the structure
around maximum light of the light-curves of some of the moderately fast (DQ
Her-like) novae in this catalogue, e.g. PACN-00-04 and PACN-00-05. However,
there may be a more marked similarity to the light-curve around maximum of the
very slow nova HR Del 1967 \citep{1977A&AS...30..323D}. We also 
note that PACN-01-01 is only 3 arcmin from the centre of M31 and hence suffers 
significantly from a highly variable background.  The points in the 1st and 2nd 
seasons likely arise from a statistical anomaly in the M31 background at or 
close to the position of this nova, and are not related to the outburst in the 
3rd season, as no resolved object is visible at this location.
 
\subsection{Distribution of classical novae}

\begin{table}
\caption{The distribution within each CCD of candidates selected by our
classical nova detection pipeline.}
\label{ccdcand2}
\begin{tabular}{lll}
\hline
CCD     & Candidates & Percentage\\
        &            & of total\\
\hline
North 1 & 4          & 20.0\% \\
North 2 & 4          & 20.0\% \\
North 3 & 1          & 5.0\%  \\
North 4 & 3          & 15.0\% \\
South 1 & 0          & 0.0\%  \\
South 2 & 0          & 0.0\%  \\
South 3 & 6          & 30.0\% \\
South 4 & 2          & 10.0\% \\
\hline
\end{tabular}
\end{table}

\begin{figure}
\rotatebox{270}{\includegraphics[width=243pt]{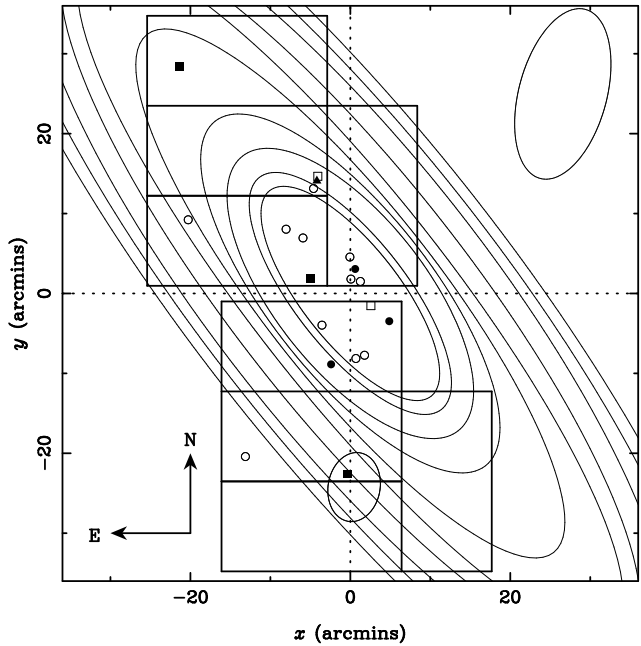}}
\caption{The positions of the selected classical nova
candidates within M31. The lines and contours are as for Figure~\ref{fields}. 
The different symbols indicate a very fast nova (solid triangle), fast novae 
(closed circles), moderately fast novae (open circles), slow novae (closed 
boxes), and very slow novae (open boxes).}\label{loc}\end{figure}

The distribution of candidates with CCD can be seen in
Table~\ref{ccdcand2}.  Figure~\ref{loc} gives a graphical
representation of each candidate's position within our fields. The
distribution shows some evidence of spatial concentration around the
bulge, however we caution that the significance of this cannot be
properly estimated before we make Monte-Carlo completeness tests of
our selection criteria, something which we will report in a follow-up
paper. One nova (PACN-00-07) lies well outside the main disk light on
the far-disk side of M31. If it is associated with the M31 disk then
it must lie at a de-projected distance of around 25~kpc from the centre
of M31, or around 4 disk scale lengths.

As can be seen from Figure \ref{loc}, one of our CNe may be
located within the dwarf spheroidal galaxy M32.  This nova,
PACN-99-04, is located 22.6 arcmin from the centre of M31, but is only
1.8 arcmin from the centre of M32.

\subsection{Pipeline detection efficiency}

In a forthcoming study we shall make a careful assessment of the efficiency with
which CNe are detected by our pipeline. However, a useful comparison can be made
with the CNe which have been announced on IAU Circulars during the lifetime of
the survey. \citet{an2004} found that 12 of the 14 CN which were
alerted during our survey are present in the POINT-AGAPE dataset. All of these
CNe are located within a few arcmins of the centre of M31 and are of fast or
moderately fast speed class. Our automated pipeline has identified 7 out of the
12 CNe listed in Table~3 of \citet{an2004}. Of the remaining five, two
(26285/26121 and 79136, using the An et al. identifiers) occur too late in the
survey to be properly sampled (and therefore failed selection). One CN (78668)
was lost due to our image trimming as it occurs close to the northern edge of
CCD3 in the southern field. Another (83479)  was lost due to the masking of a
diffraction spike of a very bright star, and the final missed CN (26277/25695)
failed the initial $15\,\sigma$ cut. We therefore conclude that our pipeline
successfully recognises CNe within the boundaries of our defined selection
criteria.

\subsection{Borderline and other light-curves} \label{borderl}

From the inspection of the CN in our catalogue, we are confident that
at least 18 of the candidates are CNe.  However
there are two CN, PACN-00-07 and PACN-01-01, that we are less
certain about.

\begin{figure}
\rotatebox{270}{\includegraphics[width=181pt]{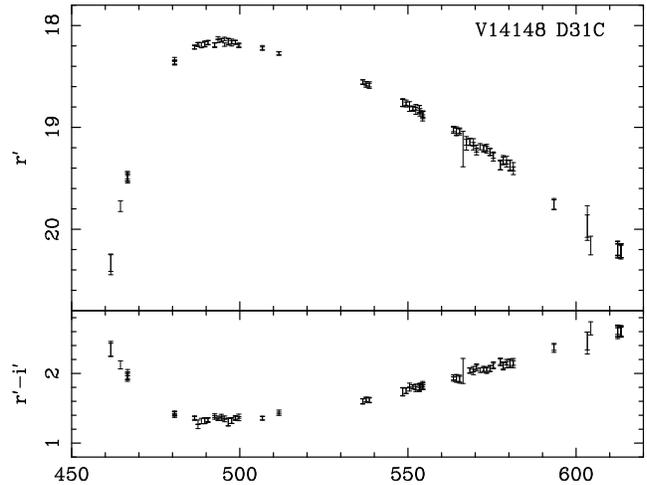}}
\caption{Mira variable V9205 D31C as observed by our survey.}
\label{mira}
\end{figure}

\begin{figure}
\rotatebox{270}{\includegraphics[width=190pt]{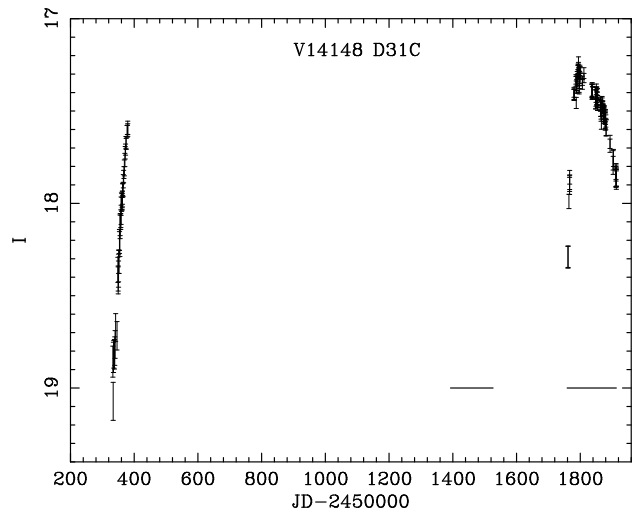}}
\caption{Mira variable V9205 D31C.  Earlier data from the DIRECT
survey \protect\citep{1999AJ....117.2810S}, later data from this
survey.  The horizontal lines indicate the length of the 1st and 2nd
POINT-AGAPE observing seasons.}
\label{mira2}
\end{figure}

 As shown in 
Table~\ref{elim}, thirty-two CN candidates were eliminated only by the two 
colour-magnitude criteria.  As can be seen in Figure~\ref{colmag}, these 
candidates appear to be located in a``clump'', indicating that they may be 
similar types of object.  From inspection, the majority of these candidates 
appear to be Miras or Mira-like variables, with a few exceptions. 
Figure~\ref{mira}, shows the light-curve of the brightest Mira discovered. Its 
position in the colour-magnitude diagram is indicated in Figure~\ref{colmag}.  
This object exhibits a very smooth light-curve and its position ($\alpha=0^{\rm 
h}44^{\rm m}23\fs7$,$\delta=+41\degr28\arcmin4\farcs1$) is coincident with an 
object exhibiting a remarkably similar light-curve from the DIRECT survey
\citep{1999AJ....117.2810S}, V9205 D31C, observed between September
and October 1996.  Figure~\ref{mira2} shows both the DIRECT $I$-band
data and POINT-AGAPE data that has been transformed into $I$-band
data, as well as the span of the 1st and 2nd POINT-AGAPE observing
seasons. The transformation of the POINT-AGAPE data to the $I$ band is
obtained by deriving a best-fit linear transformation from $i'$ and
$r'$ to $I$ using the \citet{1992A&AS...96..379M} standard stars. From a simple
analysis of the DIRECT and POINT-AGAPE data,
given that the Mira was unresolved throughout the 1st POINT-AGAPE
season, we arrive at two possible periods for this Mira, either
$\sim700$ days or $\sim1400$ days, making it one of the longest period
Miras observed. At the distance of M31, it is also one of the most luminous.

\begin{figure}
\rotatebox{270}{\includegraphics[width=190pt]{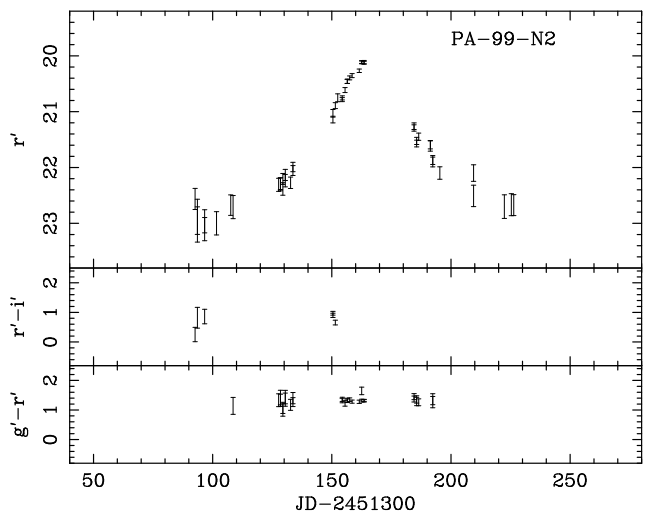}}
\caption{The PSF-fitted light-curve of the microlensing event PA-99-N2 
discovered previously by \citet{2003A&A...405...15P} using ``superpixel'' 
photometry. The event is investigated in detail by 
\citet{2004ApJ...601..845A}.}\label{PA99N2}\end{figure}

The high signal-to-noise ratio microlensing event PA-99-N2
\citep{2003A&A...405...15P} was also identified and was eliminated from the
catalogue via the colour evolution selection criterion (as would be
expected for a strong microlensing event).  The light-curve of the event 
produced using the CN detection pipeline is shown in Figure~\ref{PA99N2}. A
recent detailed analysis of this event using ``superpixel'' photometry
indicates anomalies near the peak of the light-curve which are well
explained as being due to a binary lens system \citep{2004ApJ...601..845A}. The
PSF-fitting photometry independently undertaken for this study
confirms the anomalous kink on an otherwise smooth and achromatic
light-curve, occurring on the rising side near peak.

\section{Conclusions} \label{conclusion}

The primary aim of the POINT-AGAPE survey is to search for
microlensing events due to compact dark matter in M31. However, the
requirements for such a survey also enable us to compile a substantial
catalogue of variable stars and transients. In this study we have
conducted the first fully-automated search for classical novae (CNe),
using objective selection criteria to define our sample. Our aim has
been to devise criteria which, as far as possible, do not bias against
the detection of CNe of certain speed classes in favour of others,
despite the difficulty that the CN light-curve morphology is
inextricably linked to its speed class. In the absence of H$\alpha$
observations, which are an important diagnostic for CN identification,
excellent sampling is a crucial pre-requisite for this task.

Our final catalogue of 20 CNe obtained from 3 seasons of data covering
18 months of total observing time spans a wide range of
speed class from very fast to very slow. Their light-curve morphologies
vary considerably from the smoothly declining very slow CN,
PACN-00-02, through to the moderately fast CN, PACN-00-05, which
exhibits multiple maxima as well as a deep transition minimum. Among the
objects which did not make our catalogue is V9205~D31C, first seen by
the DIRECT survey \citep{1999AJ....117.2810S}. Combining the DIRECT
data and POINT-AGAPE data reveals this object to be a Mira with a
period of either 700 or 1400 days, making it one of the longest-period
and (if as seems most likely it lies in M31) most luminous Miras known. Our 
PSF-fitting photometry also confirms the anomalous behaviour of the high
signal-to-noise ratio microlensing event PA-99-N2 \citep{2004ApJ...601..845A},
first discovered using our ``superpixel'' photometry pipeline. Overall, despite
the absence of H$\alpha$ data, we are confident that most, if not all, of our
sample are indeed CNe. The only possible borderline candidates are PACN-00-07
and PACN-01-01.

In a follow-up study we intend to use our CN catalogue to undertake an
objective study of the spatial and speed class distribution of CNe in
M31. This may help to answer key questions such as whether there is
more than one population of CN and whether CNe of different speed
classes tend to be associated with different stellar populations. From
Monte-Carlo completeness tests we should be able to assess objectively
the underlying global nova rate, the spatial distribution of CNe, and
their potential as standard candle distance estimators via the
maximum-magnitude versus rate-of-decline
and other relationships. The detailed profiles observed for several of the 
novae,especially prior to maximum, may also prove helpful for constraining
theoretical models of the nova outburst.

\section*{Acknowledgements}

The work of MJD and MFB is supported by a studentship and a Senior Fellowship
respectively from the Particle Physics and Astronomy Research Council. JA is 
supported by a grant from the Leverhulme Trust Foundation and NWE thanks the
Royal Society for support. SCN is supported by the Swiss National Science
Foundation and the Tomalla Foundation.

This research has made use of the NASA/IPAC Extragalactic Database (NED) which is operated by the Jet Propulsion Laboratory, California Institute of Technology, under contract with the National Aeronautics and Space Administration.

This research has made use of the SIMBAD database, operated at CDS, Strasbourg, France.

\bibliographystyle{mn2e}
\bibliography{refs}

\end{document}


\begin{landscape}

\setcounter{table}{5}
\begin{table*}
\begin{minipage}{219mm}
\caption{The POINT-AGAPE classical nova catalogue. $x$ and $y$ are the
positions from the M31 centre measured in arcminutes along the M31
major and minor axis, respectively.  The epoch t$_{0}(r')$ is the time
of the brightest $r'$-band observation.  The maximum magnitudes
$r'($t$_{0})$, $i'($t$_{0})$ and $g'($t$_{0})$ are the brightest observed
magnitude in each band. $dr'/dt$ is an estimate of the rate of decline
between the brightest observation and the observation closest to 2
magnitudes fainter than the maximum observed. The speed class is
estimated by assuming the $V$-band definition in Table~2 also holds
for our $r'$-band data. }
\label{canddata}
\begin{tabular}{lllllllllll}
\hline
Nova             & R.A.            & Decl.                        & $x$  & $y$  & t$_{0}$($r'$) & $r'$(t$_{0}$)          & $i'$(t$_{0}$)   & $g'$(t$_{0}$)   & $dr'/dt$          & Speed \\
                 & (J2000)         & (J2000)                      & (arcmin) & (arcmin) & JD-2451000   &                       &                &                & (mag day$^{-1}$) & class \\
                 &                 &                              &          &          &              &                       &                &                & {\em (estimate)} \\
\hline
PACN-99-01       & $0^h43^m27\fs2$ & $+41\degr24\arcmin11\farcs1$ & -8.0     & 8.0      & 392.63             & $16.53\pm0.03$  & $16.39\pm0.03$ & $17.40\pm0.01$ & 0.06        & Moderately Fast\\
PACN-99-02       & $0^h44^m39\fs2$ & $+41\degr44\arcmin32\farcs4$ & -21.4    & 28.4     & 394.65             & $18.91\pm0.03$  & $19.19\pm0.04$ & $20.22\pm0.02$ & 0.02        & Slow\\
PACN-99-03       & $0^h42^m34\fs9$ & $+41\degr08\arcmin24\farcs5$ & 1.8      & -7.7      & 395.71             & $17.79\pm0.02$  & $17.60\pm0.04$ & -              & 0.03        & Moderately Fast\\
PACN-99-04       & $0^h42^m46\fs1$ & $+40\degr53\arcmin35\farcs3$ & -0.3     & -22.6    & 400.61$^{*}$       & $18.41\pm0.02$  & $18.34\pm0.07$   & $18.83\pm0.02$ & 0.02      & Slow\\
PACN-99-05       & $0^h42^m41\fs1$ & $+41\degr19\arcmin12\farcs2$ & 0.6      & 3.1      & 427.69             & $17.70\pm0.04$  & -              & $18.47\pm0.02$ & 0.02        & Fast\\
PACN-99-06       & $0^h43^m15\fs9$ & $+41\degr23\arcmin05\farcs4$ & -5.9     & 6.9      & 432.69$^{*}$       & $16.17\pm0.01$  & -              & $16.91\pm0.01$ & 0.06        & Moderately Fast\\
PACN-99-07       & $0^h43^m06\fs7$ & $+41\degr30\arcmin14\farcs2$ & -4.2     & 14.1     & 484.50             & $18.1\pm0.1$    & $18.02\pm0.04$ & $18.5\pm0.1$ & 0.2         & Very Fast\\
PACN-00-01       & $0^h42^m44\fs0$ & $+41\degr17\arcmin56\farcs5$ & 0.1      & 1.8      & 760.52             & $17.73\pm0.04$  & $17.58\pm0.8$   & -              & 0.05         & Moderately Fast\\
PACN-00-02       & $0^h43^m06\fs0$ & $+41\degr30\arcmin48\farcs3$ & -4.1     & 14.7     & 761.63             & $18.15\pm0.03$  & $18.86\pm0.05$ & -              & 0.01        & Very Slow\\
PACN-00-03       & $0^h42^m44\fs6$ & $+41\degr20\arcmin42\farcs1$ & -0.1     & 4.6      & 766.64             & $18.54\pm0.03$  & $18.19\pm0.04$ & -              & 0.06         & Moderately Fast\\
PACN-00-04       & $0^h42^m37\fs6$ & $+41\degr17\arcmin38\farcs6$ & 1.3      & 1.5      & 766.65$^{*}$       & $17.61\pm0.03$  & $17.33\pm0.04$ & -              & 0.07        & Moderately Fast\\
PACN-00-05       & $0^h43^m08\fs9$ & $+41\degr29\arcmin16\farcs6$ & -4.6     & 13.1     & 786.54             & $17.30\pm0.01$  & $17.11\pm0.01$ & -              & 0.03        & Moderately Fast\\
PACN-00-06$^{a}$ & $0^h42^m57\fs1$ & $+41\degr07\arcmin16\farcs3$ & -2.4     & -8.9     & 838.52$^{*}$       & $17.09\pm0.01$  & $16.64\pm0.01$ & -              & 0.1         & Fast\\
PACN-00-07       & $0^h43^m53\fs8$ & $+40\degr55\arcmin43\farcs5$ & -13.1    & -20.4    & 854.46$^{*}$       & $19.53\pm0.04$  & $19.48\pm0.05$ & -              & 0.03        & Moderately Fast\\
PACN-01-01       & $0^h42^m30\fs6$ & $+41\degr14\arcmin36\farcs8$ & 2.6      & -1.5     & 1135.64            & $18.45\pm0.02$  & $18.16\pm0.04$ & -              & 0.01        & Very Slow
\\
PACN-01-02       & $0^h42^m18\fs4$ & $+41\degr12\arcmin40\farcs3$ & 4.9      & -3.5     & 1142.71$^{*}$      & $17.14\pm0.03$  & $16.71\pm0.04$ & -              & 0.09         & Fast\\
PACN-01-03       & $0^h43^m10\fs6$ & $+41\degr17\arcmin57\farcs6$ & -4.9     & 1.8      & 1148.67$^{*}$      & $17.30\pm0.04$  & $16.88\pm0.06$ & -              & 0.01        & Slow\\
PACN-01-04       & $0^h42^m40\fs6$ & $+41\degr07\arcmin59\farcs9$ & 0.7      & -8.1     & 1148.69$^{*}$      & $17.90\pm0.03$  & $17.38\pm0.04$ & -              & 0.05        & Moderately Fast\\
PACN-01-05       & $0^h44^m32\fs5$ & $+41\degr25\arcmin21\farcs9$ & -20.3    & 9.2      & 1191.48            & $15.90\pm0.01$  & $15.61\pm0.01$ & -              & 0.07        & Moderately Fast\\
PACN-01-06$^{b}$ & $0^h43^m03\fs2$ & $+41\degr12\arcmin10\farcs5$ & -3.5     & -4.0     & 1194.62$^{*}$      & $17.38\pm0.01$  & $16.88\pm0.03$ & -              & 0.04        & Moderately Fast\\
\hline
\end{tabular}

\medskip
$^{a}$ CN NMS-1 (Joshi et al. 2003).

$^{b}$ CN NMS-2 (Joshi et al. 2003).

$^{*}$ these novae have been observed before maximum-light, in their final rise phase. 

\end{minipage}
\end{table*}

\end{landscape}